\documentclass[10pt,twocolumn,letterpaper]{article}
\usepackage[utf8]{inputenc}
\usepackage[margin=0.8in]{geometry}
\usepackage{setspace}
\usepackage{graphicx}
\usepackage{booktabs}
\usepackage{amsmath}%
\usepackage[labelfont=bf,skip=5pt,justification=justified, format=plain]{caption}
\usepackage[caption=false]{subfig}
\usepackage[style=phys,articletitle=true,maxbibnames=30]{biblatex}
\usepackage [english]{babel}
\usepackage [autostyle, english = american]{csquotes}
\MakeOuterQuote{"}
\usepackage{lipsum}
\usepackage{url}

\begin{document}

\begin{singlespace}

\title{Vacancy Tuned Magnetism in LaMn$_x$Sb$_2$}
\author{Tyler J. Slade$^{1,2}$, Aashish Sapkota$^{1,2}$, John M. Wilde$^{1,2}$, Qiang Zhang$^3$, Lin-Lin Wang$^{1}$,\\ Saul H. Lapidus$^4$, Juan Schmidt$^{1,2}$, Thomas Heitmann$^5$, Sergey L. Bud’ko$^{1,2}$, Paul C. Canfield$^{1,2}$}
\date{}

\twocolumn[
\begin{@twocolumnfalse}

\maketitle

\begin{center} 
    
\textit{$^{1}$Ames National Laboratory, US DOE, Iowa State University, Ames, Iowa 50011, USA} \\ 
\textit{$^{2}$Department of Physics and Astronomy, Iowa State University, Ames, Iowa 50011, USA} \\
\textit{$^{3}$Oak Ridge National Laboratory, Oak Ridge, Tennessee 37831, USA} \\
\textit{$^{4}$X-ray Science Division, Advanced Photon Source, Argonne National Laboratory, 9700 S. Cass Ave, Argonne, Illinois 60439, USA} \\
\textit{$^{5}$The Missouri Research Reactor and Department of Physics and Astronomy, University of Missouri, Columbia, Missouri 65211, USA}

\begin{abstract}

The layered \textit{ATMPn}$_2$ (\textit{A} = Alkali earth or rare earth atom, \textit{TM} = transition metal, \textit{Pn} = Sb, Bi) compounds are widely studied for their rich magnetism and electronic structure topology. We provide a detailed characterization of the magnetic and transport properties of LaMn$_x$Sb$_2$, an understudied member of the \textit{ATMPn}$_2$ family. LaMn$_x$Sb$_2$ forms with intrinsic Mn vacancies, and we demonstrate that by varying the starting ratio of La, Mn, and Sb, we can synthetically control the Mn occupancy and produce single crystals with \textit{x} = 0.74--0.97. Magnetization and transport measurements indicate LaMn$_x$Sb$_2$ has a rich temperature-composition (\textit{T}--\textit{x}) magnetic phase diagram with physical properties strongly influenced by the Mn occupancy. LaMn$_x$Sb$_2$ orders antiferromagnetically at \textit{T}$_{1}$ = 130--180 K, where \textit{T}$_{1}$ increases with \textit{x}. Below \textit{T}$_{1}$, the \textit{T}--\textit{x} phase diagram is complicated. At high \textit{x}, there is a second transition \textit{T}$_2$ that decreases in temperature as \textit{x} is lowered, vanishing below \textit{x} $\leq$ 0.85. A third, first-order, transition \textit{T}$_3$ is detected at \textit{x} $\approx$ 0.92, and the transition temperature increases as \textit{x} is lowered, crossing above \textit{T}$_2$ near \textit{x} $\approx$ 0.9. On moving below \textit{x} $<$ 0.79, we find the crystal structure changes from the \textit{P}4/\textit{nmm} arrangement to a \textit{I}$\bar{4}$2\textit{m} structure with partially ordered Mn vacancies. The change in crystal structure results in the sudden appearance of two new low temperature phases and a crossover between regimes of negative and positive magnetoresistance when \textit{x} $\leq$ 0.78. Finally, we provide powder neutron diffraction for \textit{x} = 0.93, and find that the high-\textit{x} compositions first adopt a \textit{G}-type antiferromagnetic structure with the Mn moments aligned within the \textit{ab}-plane which is followed upon further cooling by a second transition to a different, non-collinear structure where the moments are rotated within the basal plane. Our results demonstrate that LaMn$_x$Sb$_2$ is a highly tunable material with six unique magnetically ordered phases, depending on \textit{T} and \textit{x}. 

\end{abstract}

\end{center}

\vspace{7mm}

\end{@twocolumnfalse}
]

\section{Introduction}

Defect chemistry strongly influences nearly all material properties, making study of defects in different chemical compounds necessary at a fundamental and applied level, both to understand basic physical properties and to engineer functional materials with well controlled behavior. The role of defects is particularly salient in compounds with large widths of formation, where intrinsic vacancies and/or interstitial atoms substantially shift the chemical composition from the ideal stoicheometry. In such materials, the disorder inherently produced alongside vacancies or interstitials can be undesirable, yet the flexible composition can in many cases be exploited as an intrinsic tuning parameter and used to manipulate the Fermi level, electronic structure, and the associated physical properties. In this regard, proper defect engineering is essential to achieving the desired behavior in materials and applications spanning superconductivity,\autocite{cava1987oxygen,PhysRevB.83.144517,PhysRevB.85.224528} thermoelectricity,\autocite{D1CS00347J,PRXEnergy.1.022001,OHNO2018141} and catalysis,\autocite{https://doi.org/10.1002/adfm.202009070,https://doi.org/10.1002/aenm.201600436,BAI2018296} amongst others. 

\begin{figure}[!t]
    \centering
    \includegraphics[width=\linewidth]{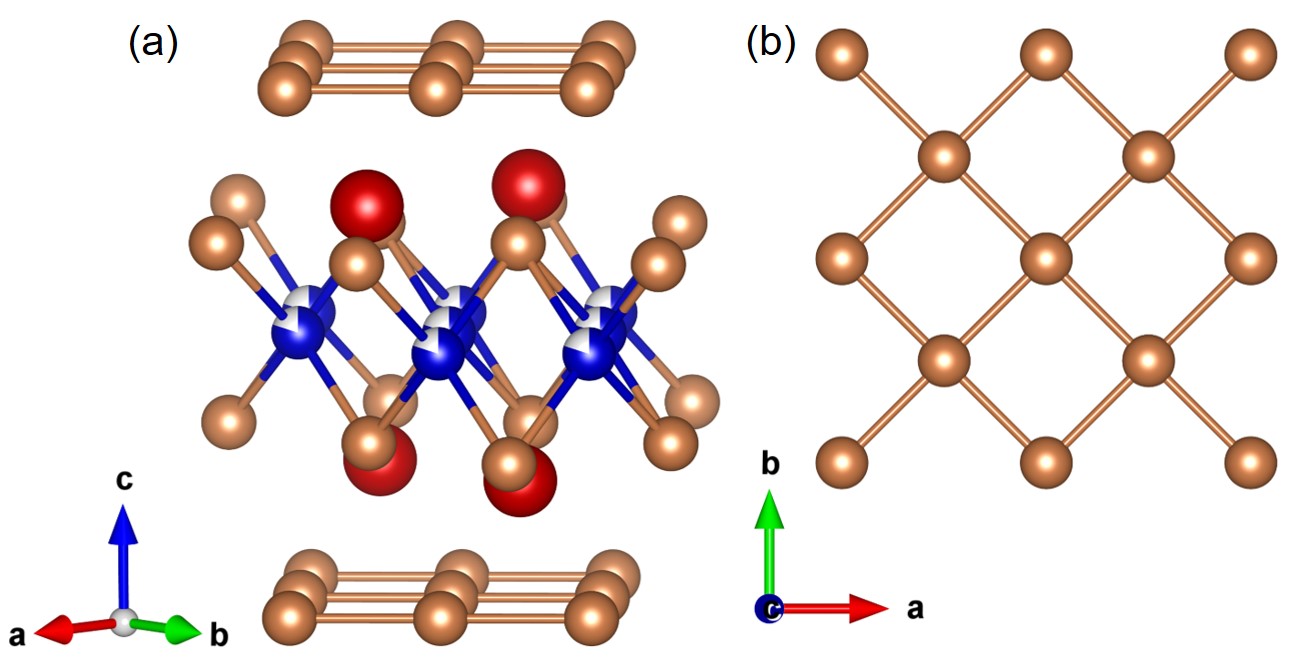}
    \caption[]{The \textit{P}4/\textit{nmm} crystal structure of LaMn$_x$Sb$_2$. (a) gives a perspective emphasizing the layered structure consisting of tetrahedrally bonded Mn-Sb sheets and square nets of Sb atoms separated by La. (b) shows an isolated square net of Sb viewed down the \textit{c}-axis. The color code is as follows: Red: La, Blue: Mn, Brass: Sb. The partial blue shading for Mn represents the partial occupancy of the Mn site (see text).}
    \label{Structure}
\end{figure}

The \textit{ATMPn}$_2$ magnetic topological semimetals, in which magnetic order and non-trivial electronic structures can coexist, are an attractive family to study the influence of vacancies or defects on their magnetism and electronic properties, and consequentially on their topological features. The \textit{RETM}$_x$Sb$_2$ compounds (\textit{TM} = transition metal) are part of the wider \textit{ATMPn}$_2$ class, in which \textit{A} is an alkali-earth or rare-earth (RE) element, and Pn is Sb or Bi. Depending on the identity of the \textit{A} and Pn atoms, the \textit{ATMPn}$_2$ materials adopt a layered tetragonal structure (see the example of LaMn$_x$Sb$_2$ shown in Figure \ref{Structure}), or lower symmetry distortion of the parent structure. A prominent structural motif common to these materials is a square net of Pn atoms that is separated from the tetrahedrally bonded \textit{T}--\textit{Pn} layers by the \textit{A} ions. The symmetry of the square net protects linear crossings (Dirac points) of the bands derived from the pnictogen \textit{p$_x$} and \textit{p$_y$} orbitals,\autocite{klemenz2019topological,klemenz2020role,hyun2018unified} providing topological physics if the Fermi level is near the Dirac points. Because the rare earth and transition metal sites can host host moment bearing atoms, the \textit{ATMPn}$_2$ materials have been exhaustively studied as magnetic topological semimetals.\autocite{feng2014strong,park2011anisotropic,wang2011layered,he2012giant,wang2012two,liu2017unusual,borisenko2019time,wang2018quantum,kealhofer2018observation,yi2017large,soh2019magnetic,masuda2016quantum,zhu2020magnetic,PhysRevB.100.195123,10.1063/1.4758298,PhysRevB.97.115166,chamorro2019dirac,zhang2022toward,yi2017large,liu2017magnetic,PhysRevB.100.014437}  

Most existing work on the \textit{ATMPn}$_2$ family focuses on members in which \textit{A} is a divalent cation, normally an akali earth element, Eu, or Yb. On the other hand, most of the \textit{ATMPn}$_2$ compounds in which \textit{A} is a trivalent rare earth element are yet to be widely studied. This may be part be due to their large widths of formation. With the exception of the case in which \textit{TM} = Ag,\autocite{myers1999systematic} most of the \textit{RE}\textit{TM}Sb$_2$ materials with trivalent \textit{RE} (La-Nd, Sm) form with a high fraction of transition metal vacancies and have true compositions of \textit{RETM}$_x$Sb$_2$ with \textit{x} $\approx$ 0.5-0.9.\autocite{sologub1995ternary,leithe1994crystal,wollesen1996ternary} The vacancies introduce undesired disorder, which may obscure the quantum transport properties associated with the symmetry protected energy bands. Nevertheless, vacancies will naturally hole dope the material, so if the Mn occupancy can be chemically manipulated, this may be a feasible means to control the physical properties, as was observed in the isostructural LaAu$_x$Sb$_2$, where altering the Au vacancy density between \textit{x} =0.9--1 allowed for tuning of the charge density wave transition temperatures by up to 80 K.\autocite{PhysRevB.102.125110}

Here, we report the compositional dependence of the magnetic phases of LaMn$_x$Sb$_2$. The limited data available on LaMn$_x$Sb$_2$ report conflicting magnetic behavior. Initially, work by Sologub et al. on several polycrystalline samples with \textit{x} $\approx$ 0.65--0.9 indicated ferromagnetic order below T$_C$ $\approx$ 350 K.\autocite{sologub1995ternary} More recently, Li et al. analyzed single crystals of LaMn$_x$Sb$_2$ and found evidence for an initial ferromagnetic or spin-canted antiferromagnetic, transition at 138 K followed by a second, antiferromagnetic transition at 56 K, but no evidence for magnetic order near or above room temperature.\autocite{PhysRevB.105.224429} Likewise, Yang et al. observed a single antiferromagnetic transition \textit{T}$_N$ = 146 K and 138 K in single crystals with respective compositions LaMn$_{0.86}$Sb$_2$ and LaMn$_{0.84}$Sb$_2$.\autocite{PhysRevB.107.115150} Whereas the recent work on single crystals is limited to only several compositions, \textit{x} = 0.84, 0.86, and 0.91, the observance of different magnetic states and ordering temperatures for different compositions supports the notion that controlling the Mn vacancies may allow for tuning of the magnetic order.

To determine what magnetic states are found in LaMn$_x$Sb$_2$ and clarify how these depend on the Mn occupancy, we conducted an exhaustive characterization of LaMn$_x$Sb$_2$ single crystals. We grow single crystals of LaMn$_x$Sb$_2$ from a ternary La-Mn-Sb solution, and by changing the starting melt composition, we are able to synthetically control the Mn occupancy to produce crystals with \textit{x} = 0.74--0.97. In contrast to the earlier reports of ferromagnetic order, we find that LaMn$_x$Sb$_2$ orders antiferromagnetically below \textit{T}$_{1}$ = 130--180 K, depending on the Mn occupancy, and we ascribe the apparent room temperature ferromagnetic behavior in the older literature, pertaining to polycrystalline samples, to small amounts of a ferromagnetic impurity, MnSb. We find the antiferromagnetism in LaMn$_x$Sb$_2$ to be complex and highly sensitive to the Mn stoichiometry, and provide evidence for six different magnetic states in LaMn$_x$Sb$_2$ depending on the temperature and \textit{x}. Field dependent magnetization data points to an equally complex temperature-field phase diagram when \textit{x} $>$ 0.93. Powder neutron diffraction measurements reveal that when \textit{x} $\geq$ 0.93, LaMn$_x$Sb$_2$ initially adopts a \textit{G}-type antiferromagnetic structure below 170 K in which the Mn moments are oriented in the \textit{ab}-plane. On cooling below 150 K, we find a different non-collinear arrangement where the Mn moments are rotated within the basal plane. We furthermore find evidence that the crystal structure changes from \textit{P}4/\textit{nmm} when \textit{x} $\geq$ 0.78 to a different, \textit{I}$\bar{4}$2\textit{m}, structure when \textit{x} $\leq$ 0.78, associated with partial ordering of the Mn vacancies. Given the ability to tune between six magnetically ordered states and two crystal structures, our work ultimately shows that LaMn$_x$Sb$_2$ is an extremely sensitive material and may serve as an excellent model system for studying the effects of vacancies and disorder.


\section{Experimental Methods}

\begin{figure*}[!t]
    \centering
    \includegraphics[width=\linewidth]{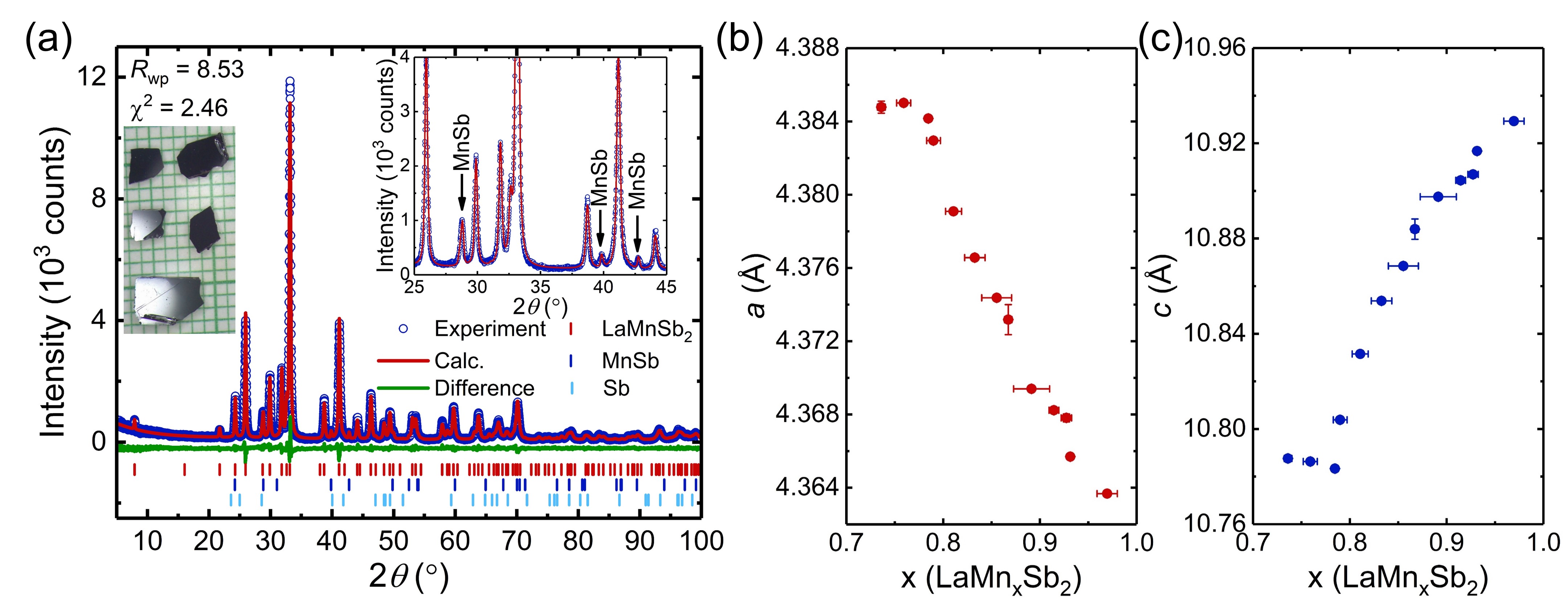}
    \caption[]{(a) Powder X-ray diffraction pattern from LaMn$_x$Sb$_2$ with \textit{x} = 0.93. The solid lines are the calculated pattern from a Rietveld refinement using the \textit{P}4/\textit{nmm} structural model. The right inset shows a closeup of several small reflections from MnSb (marked with arrows), and the left inset shows typical crystals on a mm grid. (b) and (c) respectively show the Rietveld refined lattice parameters \textit{a} and\textit{c} as a function of Mn occupancy \textit{x} for LaMn$_x$Sb$_2$. The horizontal error bars were determined from the deviation in the values of \textit{x} from the PXRD refinements and EDS measurements.}
    \label{pxrd}
\end{figure*}

\textbf{\textit{Crystal growth}}: Single crystals of LaMn$_x$Sb$_2$ were grown from ternary La-Mn-Sb melts. Because LaMn$_x$Sb$_2$ has a known width of formation, we attempted to grow crystals with different Mn occupancy (\textit{x}) by varying the stoichiometry of the starting mixture. To grow the crystals, elemental La, Mn, and Sb were weighed (according to the compositions in Table \ref{x_table}) and placed into 2 mL frit-disc alumina crucible sets (Canfield crucible sets, CCS, sold by LSP Ceramics)\autocite{canfield2016use,lspceramics_CCS} and flame sealed under vacuum in fused silica ampules. The tubes were loaded into a box furnace and brought to 1150°C over 12 h and held at temperature for 6 h. The furnace was then cooled over $\approx$ 150 h to 850°C after which the tubes were quickly removed from the furnace and the excess flux decanted using a centrifuge with a specially fabricated metal rotor and cups.\autocite{canfield2019new} After cooling to room temperature, the tubes were broken open to reveal metallic plates up to 1 cm in length. Pictures of typical crystals are shown in the left inset to Figure \ref{pxrd}a. 

We emphasize that after cooling, the decanted material (solidified flux captured in the catch crucible) was strongly attracted to a Nd$_2$Fe$_{14}$B magnet, indicating a ferromagnetic material forms from the excess liquid phase. In agreement with this expectation, powder X-ray diffraction patterns obtained on the solidified decant indicate a significant fraction of the binary compound MnSb (see Figure \ref{SpinPXRD} below in the Appendix), which is a known ferromagnet with \textit{T}$_C$ $>$ 300 K.\autocite{okita1968crystal,teramoto1968existence} Furthermore, small droplets/patches of solidified flux that were present on the surfaces of our crystals were also attracted to a magnet, whereas the bulk crystals were not (after carefully removing all droplets and polishing the surfaces).

\begin{table}[t!]
  \centering
  \caption{Nominal La$_a$Mn$_b$Sb$_c$ compositions used to grow LaMn$_x$Sb$_2$ and values of the Mn occupancy (\textit{x} estimated from Rietveld refinements of the PXRD patterns and energy dispersive spectroscopy (EDS).}
    \begin{tabular}{lrr}
    \toprule
    Melt Composition & \multicolumn{1}{l}{x (refined)} & \multicolumn{1}{l}{x (EDS)} \\
    \midrule
    La$_{15}$Mn$_{22}$Sb$_{63}$ & 0.739(3) & 0.73(3) \\
    La$_{15}$Mn$_{25}$Sb$_{65}$ & 0.764(3) & 0.75(3) \\
    La$_{19}$Mn$_{19}$Sb$_{62}$ & 0.785(3) & - \\
    La$_{7}$Mn$_{29}$Sb$_{64}$ & 0.795(1) & 0.78(2) \\
    La$_{7}$Mn$_{36}$Sb$_{57}$ & 0.803(4) & 0.82(2) \\
    La$_7$Mn$_{40}$Sb$_{53}$ & 0.825(4) & 0.84(3) \\
    La$_{15}$Mn$_{35}$Sb$_{50}$ & 0.86(1) & 0.84(3) \\
    La$_7$Mn$_{44}$Sb$_{49}$ & 0.87(3) & - \\
    La$_7$Mn$_{46}$Sb$_{47}$ & 0.904(5) & 0.87(2) \\
    La$_7$Mn$_{47}$Sb$_{46}$ & 0.92(2) & 0.91(2) \\
    La$_7$Mn$_{47}$Sb$_{46}$ & 0.920(2) & 0.93(2) \\
    La$_7$Mn$_{48}$Sb$_{45}$ & 0.932(5) & 0.91(3) \\
    La$_7$Mn$_{50}$Sb$_{43}$ & 0.959(5) & 0.98(2) \\
    \bottomrule
    \end{tabular}%
  \label{x_table}%
\end{table}%

\noindent
\textbf{\textit{Powder X-ray diffraction}}: The identity of the crystals were determined by powder X-ray diffraction (PXRD). Several crystals from each batch (corresponding to the compositions in Table \ref{x_table}) were ground to a powder, sifted through a 45 micron mesh sieve. In-house (laboratory) diffraction patterns were collected at room temperature on a Rigaku Miniflex-II instrument operating with Cu-K$\alpha$ radiation, $\lambda$ = 1.5406 Å (K$\alpha_1$) and 1.5443 (K$\alpha_2$) Å, at 30 kV and 15 mA. High resolution PXRD was performed on samples with \textit{x} = 0.93 and \textit{x} = 0.74 at 295 K and 90 K at the Advanced Photon Source 11-BM at Argonne National Laboratory and with a wavelength of $\lambda$ = 0.458949 Å. Rietveld refinements of the experimental powder patterns were used to determine the unit cell parameters and estimate the Mn site occupancy (\textit{x}). The powder data was refined using GSAS-II software.\autocite{toby2013gsas} To ensure reasonable reproducibility and estimate the uncertainty, we collected and refined two or three separate patterns from each batch of crystals. The crystal structures were visualized using VESTA software.\autocite{momma2011vesta}

\noindent
\textbf{\textit{Energy Dispersive Spectroscopy (EDS)}}: The Mn occupancy of the LaMn$_x$Sb$_2$ samples were also inferred using energy dispersive spectroscopy (EDS) quantitative chemical analysis using an EDS detector (Thermo NORAN Microanalysis System, model C10001) attached to a JEOL scanning-electron microscope (SEM). The compositions of the crystals were measured at ten different positions on the crystal's face (perpendicular to \textit{c} axis), revealing good homogeneity in each crystal. An acceleration voltage of 11 kV, working distance of 10 mm and take off angle of 35 deg were used for measuring all standards and crystals with unknown composition. A LaSb$_2$ single crystal was used as a standard for La and Sb quantification,\autocite{PhysRevB.57.13624} and a SmMn$_2$Ge$_2$ single crystal was used as a standard for Mn.\autocite{PhysRevB.62.R6073} The spectra were fitted using NIST-DTSA II Microscopium software.\autocite{ritchie2011standards} The average compositions and error bars were obtained from these data, accounting for both inhomogeneity and goodness of fit of each spectra.

\noindent
\textbf{\textit{Magnetic property measurements}}: The magnetization measurements were performed in a Quantum Design Magnetic Property Measurement System SQUID magnetometer. Prior to each measurement, any visible surface contamination of MnSb was cut away and the crystals were carefully polished on all surfaces until they were no longer attracted to a Nd$_2$Fe$_{14}$B magnet at room temperature (see the crystal growth section and the appendix for more details). The measurements were conducted with the field oriented parallel and perpendicular to the \textit{c}-axis, where \textit{c} is perpendicular to the plate-like surface of the single crystals. The samples were mounted on a plastic (Kel-F) disk, and a blank background using the bare disk was first measured and the values subtracted out from the sample data. The magnetic transition temperatures were determined by the extrema in the temperature-susceptibility derivative, \textit{d($\chi$\textit{T})/d\textit{T}}.\autocite{fisher1962relation}

\noindent
\textbf{\textit{Electronic transport properties}}: The temperature and field dependent electrical resistivity was measured in a Quantum Design Physical Property Measurement System (PPMS). The measurements were conducted in standard four point geometry. The samples were prepared by cutting the crystals into bars along the plate edges. The contacts were made by spot welding 25 $\mu$m thick annealed Pt wire onto the LaMn$_x$Sb$_2$ samples, giving contact resistance generally $\approx$ 1 $\Omega$. To ensure mechanical durability, a small amount of silver expoxy was painted on top of the spot welded contacts. In all cases, the current was applied within the \textit{ab}-plane, and the field was applied along the \textit{c}-axis for the magneoresistance measurements.

\noindent
\textbf{\textit{Neutron Diffraction}}:
To determine the magnetic structures, neutron powder diffraction measurements were performed using time-of-flight powder diffractometer, POWGEN, at Spallation Neutron Source at Oak Ridge National Laboratory. Approximately 1.6 g of samples were prepared by grinding several crystals and passing the powder through a 45 micron sieve. Diffraction patterns were collected at 200, 160 and 20 K using neutron beam with center wavelengths of 1.5 and 2.665 \AA. 

Single-crystal neutron diffraction were measured on the TRIAX triple-axis-neutron spectrometer at the University of Missouri Research Reactor. Measurements on TRIAX were made using a neutron wavelength of $\lambda$ = 2.359 \AA\ selected by a pyrolytic-graphite (PG) monochromator.

\noindent
\textbf{\textit{Computational methods}}: Band structure and total energy for LaMn$_x$Sb$_2$ (assuming a disorder free, \textit{x} = 1 composition) were calculated in density functional theory\autocite{PhysRev.136.B864,PhysRev.140.A1133} (DFT) using PBE\autocite{PhysRevLett.77.3865} as exchange-correlation functional with spin-orbit coupling (SOC) included. All DFT calculations were performed in VASP\autocite{PhysRevB.54.11169,KRESSE199615} with a plane-wave basis set and projector augmented wave\autocite{PhysRevB.50.17953} method. We used the tetragonal unit cells with a $\Gamma$-centered Monkhorst-Pack\autocite{PhysRevB.13.5188} (11$\times$11$\times$4) \textit{k}-point mesh with a Gaussian smearing of 0.05 eV. The kinetic energy cutoff was 270 eV. 

\section{Results}

\begin{figure*}[!t]
    \centering
    \includegraphics[width=\linewidth]{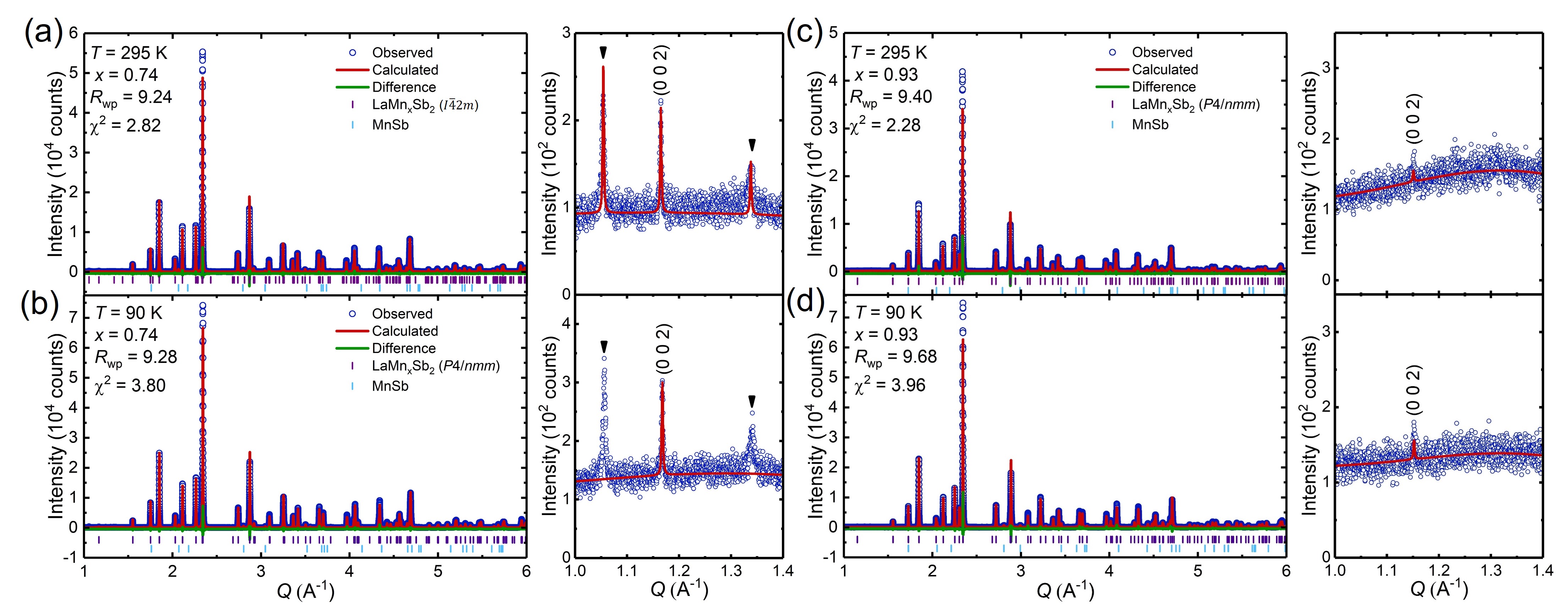}
    \caption[]{Synchrotron powder X-ray diffraction patterns for LaMn$_x$Sb$_2$ samples collected at 295 K and 90 K. (a) and (b) show the data for \textit{x} = 0.74 and (c) and (d) the data for \textit{x} = 0.93. The left panels show the full patterns and the right panels zoomed views of the low-\textit{Q} data. The solid red line in each panel shows the theoretical pattern based on the Rietveld refinements in \textit{P}4/\textit{nmm} or \textit{I}$\bar{4}$2\textit{m} as listed in the legends. The data in (b) is shown with a refinement in the \textit{P}4/\textit{nmm} arrangement to emphasize the two low-\textit{Q} peaks (marked with black arrows) not captured by this structural model.}
    \label{SynPXRD}
\end{figure*}

\subsection{Composition and phase analysis} Figure \ref{pxrd}a shows a typical powder XRD pattern obtained from finely ground \textit{x} = 0.93 crystals (details on the determination of \textit{x} in the following discussion). Similar patterns were obtained from crystals from each of the different batches listed in Table \ref{x_table}, and the primary reflections observed in the powder data are in excellent agreement with the \textit{P}4/\textit{nmm} structure expected of LaMn$_x$Sb$_2$, confirming growth of the desired phase. Likewise, Rietveld refinements of the powder patterns using the \textit{P}4/\textit{nmm} structural model give reasonable refinement statistics with \textit{R}$_{\text{wp}}$ $<$ 10 and GOF $<$ 2 for all cases. 

The powder patterns for all samples generally show weak reflections corresponding to MnSb (marked by arrows in the inset of Figure \ref{pxrd}a). PXRD analysis indicates that MnSb readily forms out of the decanted liquid phase (Figure \ref{SpinPXRD} in the Appendix), implying that the MnSb detected in Figure \ref{pxrd}b likely is from residual flux which remains on the surfaces of the crystals after decanting. Supporting this interpretation, the as-grown crystals normally have small patches/droplets of solidified flux on their surfaces that are easily visible upon inspection with a microscope. We found different decanting temperatures to have a negligible impact on the amount of MnSb left on the LaMn$_x$Sb$_2$ crystals. Because MnSb is a \textit{T}$_C$ $>$ 300 K ferromagnet, we carefully polished all surfaces of the LaMn$_x$Sb$_2$ crystals prior to conducting the magnetization measurements in order to ensure the samples were sufficiently free of the undesired second phase (see the experimental section and appendix for additional details).

As described in the experimental section, we attempted to synthetically control the Mn occupancy by growing samples from melts with different La$_a$Mn$_b$Sb$_c$ stoichiometry. For each initial melt composition, Table \ref{x_table} lists the values of \textit{x} determined from Rietveld refinements of the powder diffraction data and from energy dispersive spectroscopy (EDS). The Rietveld refined and EDS results are in reasonable agreement, with both showing an increase in \textit{x} from $\approx$ 0.74--0.97 as the melt compositions become more Mn-rich. These values are within the span of \textit{x} previously reported for polycrystalline samples.\autocite{sologub1995ternary} For the remainder of this manuscript, the \textit{x} values used for each sample are the averages of the refined and EDS values, (generally from 2-3 separate refinements and 1-2 EDS measurements collected from each batch, see experimental details for more information).

Figures \ref{pxrd}b and \ref{pxrd}c show the refined lattice parameters as a function of the Mn occupancy. Between \textit{x} = 0.79--0.97, the in-plane \textit{a} lattice parameter shrinks monotonically with increasing \textit{x}, from 4.385 \AA\ to 4.365 \AA\, while the axial \textit{c} lattice parameter grows from 10.78 \AA\ to 10.92 \AA\ as \textit{x} increases. Below \textit{x} = 0.79, \textit{a} is effectively constant at $\approx$ 4.835 \AA, while \textit{c} slightly increases as \textit{x} is lowered to 0.74.

The deviation in the \textit{x} dependence of the lattice parameters below \textit{x} = 0.79 may suggest a structural change occurs when \textit{x} becomes lower than $\approx$ 0.79. Because our laboratory PXRD patterns for samples with \textit{x} $<$ 0.79 do not show any Bragg peaks that are unexpected based on the \textit{P}4/\textit{nmm} structure (or identifiable impurities, i.e. MnSb), we conducted high-resolution synchrotron PXRD data at the Advanced Photon Source. Figure \ref{SynPXRD} displays the high-resolution powder data collected at 295 K and 90 K for samples with \textit{x} = 0.74 and \textit{x} = 0.93. All reflections found in the patterns for \textit{x} = 0.93 (Figure \ref{SynPXRD}c and \ref{SynPXRD}d) are attributable to the \textit{P}4/\textit{nmm} structure (or a small MnSb impurity) and are in good agreement with the laboratory PXRD data. On the other hand, the patterns for \textit{x} = 0.74 (Figures \ref{SynPXRD}a and \ref{SynPXRD}b) show several weak, but clearly well resolved, reflections at \textit{Q} = 1.05 and 1.34. The refinement in Figure \ref{SynPXRD}b explicitly shows that the two new peaks are inconsistent with the \textit{P}4/\textit{nmm} structure, and furthermore, the peaks cannot be matched with any known binary or ternary compounds in the La-Mn-Sb phase space, suggesting they are intrinsic to the sample. The presence of new reflections strongly supports our inference from the lattice parameters, indicating the structure of LaMn$_x$Sb$_2$ changes from \textit{P}4/\textit{nmm} to a different arrangement when \textit{x} is below $\approx$ 0.79. 

The two new reflections can be indexed with a propagation vector of (1/2 1/2 1/2) where the lower \textit{Q} peak is (1/2 1/2 1/2) and the higher \textit{Q} peak is (1/2 1/2 3/2). Since the \textit{P}4/\textit{nmm} crystal symmetry captures the vast majority of the peaks, as shown in Figure \ref{SynPXRD}b, it is likely that the crystal symmetry for \textit{x} $<$ 0.79 is closely related to, or even a subgroup of \textit{P}4/\textit{nmm}. Hence, using the ISOTROPY suite,\autocite{Campbell:wf5017,Isodistort} we explored the possible subgroups of \textit{P}4/\textit{nmm} with a $\tau$-vector of (1/2 1/2 1/2). Among nine possible subgroups, three belong to tetragonal crystal systems with body-centering, and the remaining are orthorhombic. We attempted to refine the powder data using each of the possible subgroups, and found that \textit{I}$\bar{4}$2\textit{m} ($\#$ 121), with the basis (1, 1, 0) (-1, 1, 0) and (0 0 2) with respect to the \textit{P}4/\textit{nmm} space group, best agrees with the \textit{X}-ray diffraction data. The atomic positions and structural parameters for the \textit{I}$\bar{4}$2\textit{m} structural model are given in Table \ref{table_I-42m}. As shown in Figure \ref{SynPXRD}a, the Rietveld refinement captures both the low-\textit{Q} peaks with statistical parameters of \textit{R}$_{\text{wp}}$ = 9.24 and $\chi^2$ = 2.82, indicating a reasonable refinement. We emphasize that due to the low intensity of the new Bragg peaks, the final refinement statistics are not significantly different for \textit{P}4/\textit{nmm} and \textit{I}$\bar{4}$2\textit{m} structural models, so we assessed the quality of the different structural models by direct comparison with the experimental diffraction data. The other possible subgroups either failed to capture the observed peaks or produced additional peaks not observed in our powder patterns, giving reasonable confidence in the \textit{I}$\bar{4}$2\textit{m} symmetry.

\begin{figure}[!t]
    \centering
    \includegraphics[width=\linewidth]{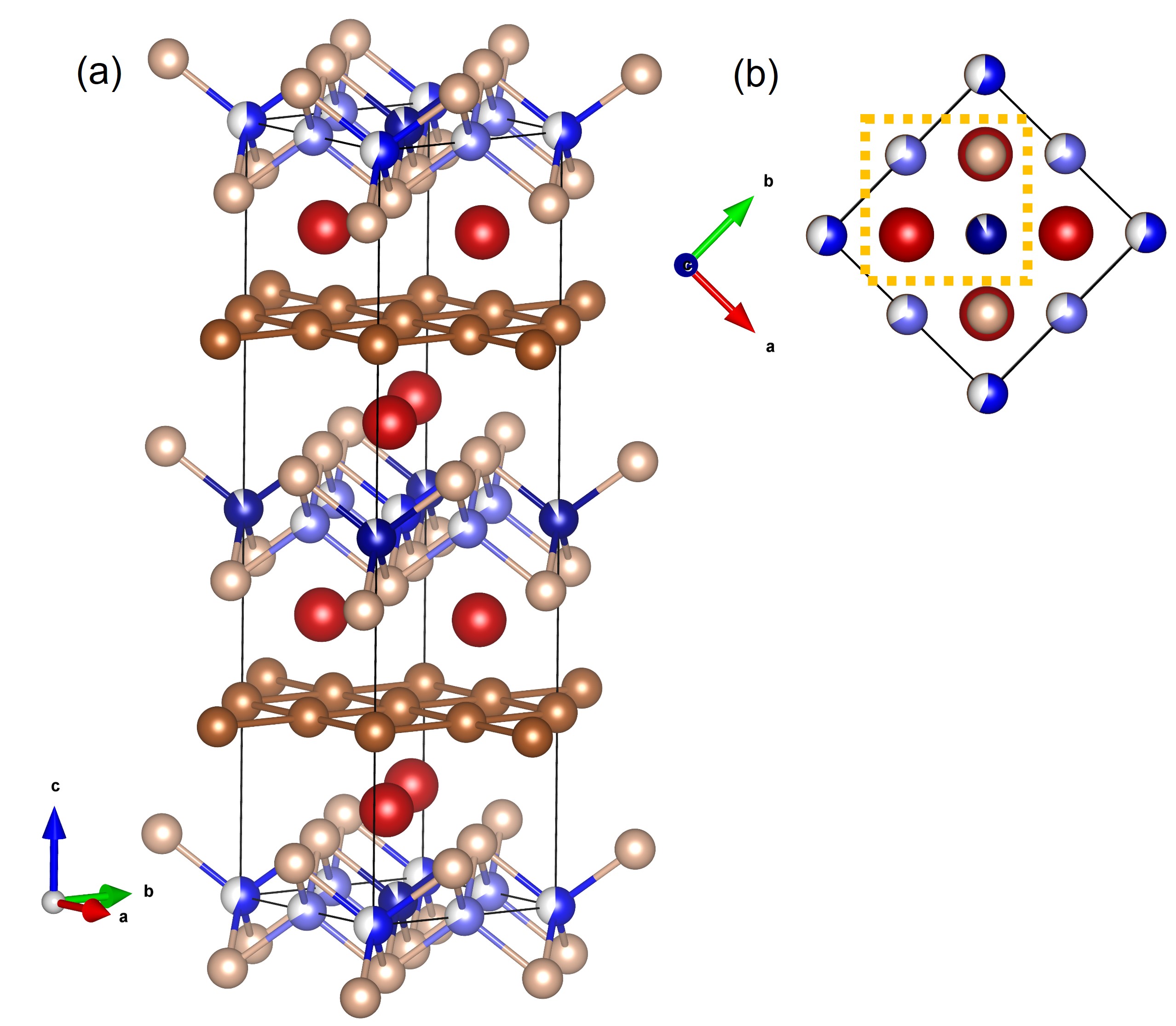}
    \caption[]{(a) \textit{I}$\bar{4}$2\textit{m} structure for LaMn$_x$Sb$_2$ with \textit{x} $<$ 0.79. (b) The same structure viewed down the \textit{c}-axis. The dashed yellow square denotes the \textit{P}4/\textit{nmm} unit cell of the \textit{x} $>$ 0.79 LaMn$_x$Sb$_2$. The color code is: red = La, blue = Mn, and brass = Sb. The partial shading for Mn atoms represnts the partial occupancy.}
    \label{structure_lowx}
\end{figure}

Figure \ref{structure_lowx} shows the \textit{x} $<$ 0.79 crystal structure. The structure is very similar to the parent \textit{P}4/\textit{nmm} arrangement, but with partial ordering of the Mn vacancies. Instead of a single Mn site, the low-\textit{x} structure has three Mn sites, with Wyckoff positions 4\textit{c}, 2\textit{a}, and 2\textit{b}. The Mn occupancy on each site is different, as illustrated in Figure \ref{structure_lowx} and listed in Table \ref{table_I-42m}, and the 2\textit{b} site is nearly fully occupied at \textit{f} $\approx$ 0.92. Compared to the high-\textit{x} \textit{P}4/\textit{nmm} structure, the partial ordering of the Mn vacancies onto three distinct crystallographic sites produces longer periodicity along both the \textit{a} and \textit{c} axis, such that the unit cell is doubled along \textit{c} (\textit{c}$_{I\bar{4}2m}$ = 2\textit{c}$_{P\text{4}/nmm}$). Likewise, the \textit{I}$\bar{4}$2\textit{m} unit cell is rotated by 45 degrees around the \textit{c} axis relative to the \textit{P}4/\textit{nmm}, and \textit{a}$_{I\bar{4}2m}$ = $\sqrt{2}$\textit{a}$_{P\text{4}/mmm}$.

\begin{table}[!t]
  \centering
  \caption{Rietveld refinement information for LaMn$_x$Sb$_2$ using the \textit{I}$\bar{4}$2\textit{m} structural model and data collected at 295 K (shown in Figure \ref{SynPXRD}a). The lattice parameters are \textit{a} = 6.20118(1) and \textit{c} = 21.57376(5), and the refinement statistics are \textit{R}$_{\text{wp}}$ = 9.24 and $\chi^2$ = 2.82.}
    \resizebox{\linewidth}{!}{\begin{tabular}{llrrrrl}
    \toprule
    label & site & \multicolumn{1}{l}{\textit{x}} & \multicolumn{1}{l}{\textit{y}} & \multicolumn{1}{l}{\textit{z}} & \multicolumn{1}{l}{frac} & \textit{U}$_{\text{iso}}$ \\
    \midrule
    La1 & 8i  & \multicolumn{1}{l}{0.2497(2)} & \multicolumn{1}{l}{0.2497(2)} & \multicolumn{1}{l}{0.36644(1)} & 1   & 0.00209(6) \\
    La2 & 8i  & \multicolumn{1}{l}{0.2508(3)} & \multicolumn{1}{l}{0.2508(3)} & \multicolumn{1}{l}{0.077170(14)} & 1   & 0.00339(8) \\
    Mn1 & 4c  & 0   & 0.5 & 0   & \multicolumn{1}{l}{0.668(13)} & 0.0027(5) \\
    Mn2 & 2a  & 0   & 0   & 0   & \multicolumn{1}{l}{0.572(15)} & \multicolumn{1}{r}{0.0102} \\
    Mn3 & 2b  & 0   & 0   & 0.5 & \multicolumn{1}{l}{0.918(15)} & \multicolumn{1}{r}{0.005} \\
    Sb1 & 4d  & 0   & 0.5 & 0.25 & 1   & \multicolumn{1}{r}{0.002} \\
    Sb2 & 4e  & 0   & 0   & 0.25 & 1   & 0.00237(13) \\
    \bottomrule
    \end{tabular}}%
  \label{table_I-42m}%
\end{table}%

\noindent
\subsection{Physical Properties} 

\begin{figure*}[!t]
    \centering
    \includegraphics[width=\linewidth]{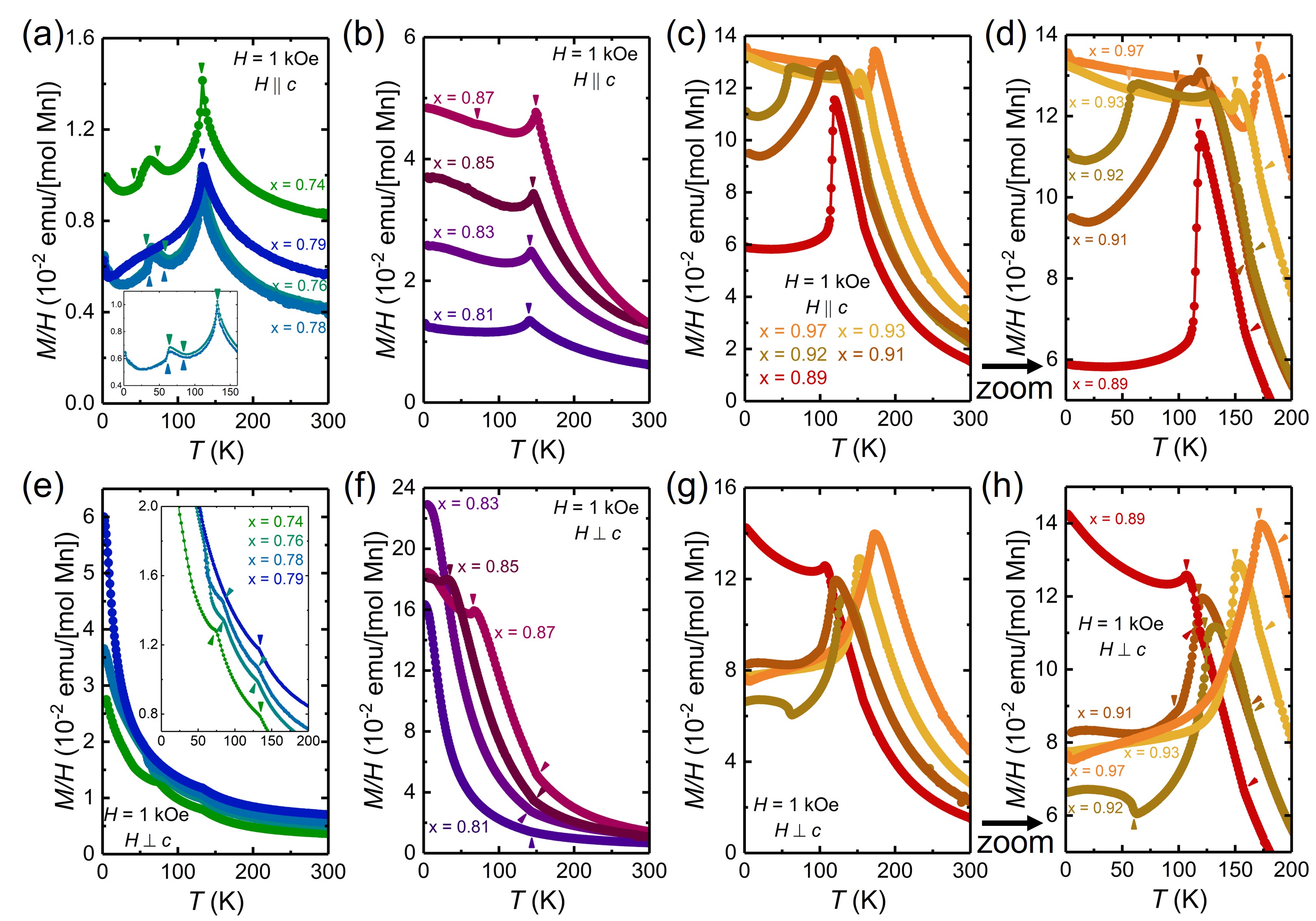}
    \caption[]{Temperature dependent magnetic susceptibility (\textit{M/H}) for LaMn$_x$Sb$_2$ (all field-cooled). The panels display data for different ranges of \textit{x} as follows: (a)--(d) show the data collected with \textit{H} $\parallel$ \textit{c} where (a) \textit{x} = 0.74--0.79, (b) \textit{x} = 0.81--0.87, and (c) \textit{x} = 0.89--0.97. (d) is a close up view of the same data in (c). (e)--(f) show data the same respective ranges of \textit{x} for \textit{H} $\perp$ \textit{c}. The insets in (a) and (e) show close up views of the data at lower temperature. The arrows mark the transitions as determined by the d($\chi$\textit{T})/d\textit{T} derivatives (see appendix). Note the different scale bars for different panels.}
    \label{MT_1kOe}
\end{figure*}

\begin{figure*}[!t]
    \centering
    \includegraphics[width=\linewidth]{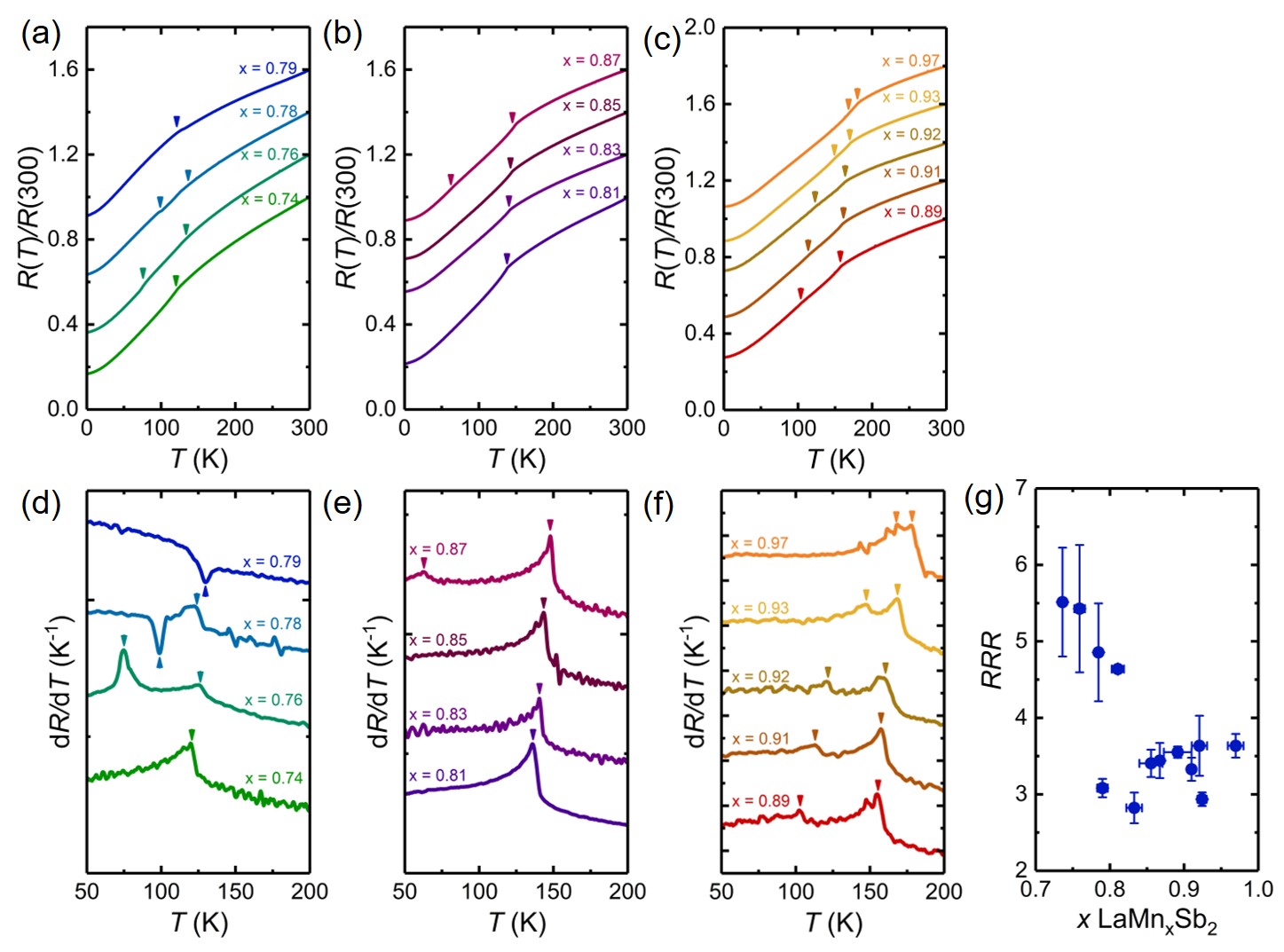}
    \caption[]{Temperature dependence of the normalized resistance \textit{R}(\textit{T})/\textit{R}(300 K) for LaMn$_x$Sb$_2$. The curves are offset for clarity. The panels show data for different ranges of \textit{x} as follows: (a) \textit{x} = 0.74--0.79, (b) \textit{x} = 0.81--0.87, and (c) \textit{x} = 0.89--0.97. (d)--(f) show the respective derivatives d\textit{R}/d\textit{T} corresponding to the data in (a)--(c). The arrows denote the magnetic transitions. (g) Residual resistance ratio (RRR) for each sample, where the RRR's are the average of 3-5 measurements for each \textit{x}.}
    \label{RT_all}
\end{figure*}

\begin{figure*}[!t]
    \centering
    \includegraphics[width=\linewidth]{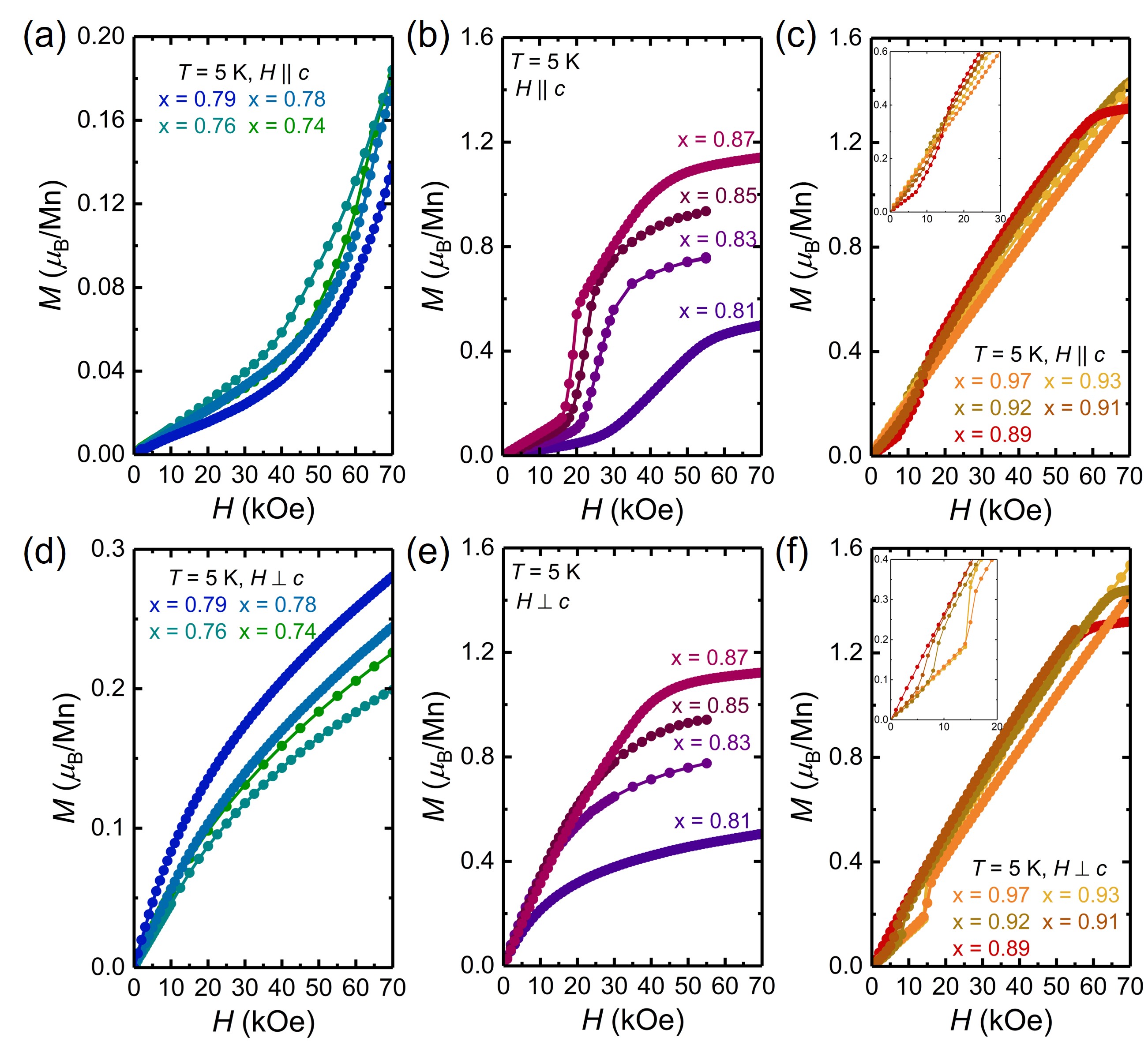}
    \caption[]{Field dependent magnetization for LaMn$_x$Sb$_2$ measured at 5 K. The pannals show data for different ranges of \textit{x} as follows: (a)--(c) display data collected with \textit{H} $\parallel$ \textit{c} for (a) \textit{x} = 0.74--0.79, (b) \textit{x} = 0.81--0.87, and (c) \textit{x} = 0.89--0.97. (d)--(f) show data the same respective ranges of \textit{x} for \textit{H} $\perp$ \textit{c}. The insets in (c) and (f) are close-up views of the low-field data.}
    \label{MH_5K}
\end{figure*}

\begin{figure}[!t]
    \centering
    \includegraphics[width=\linewidth]{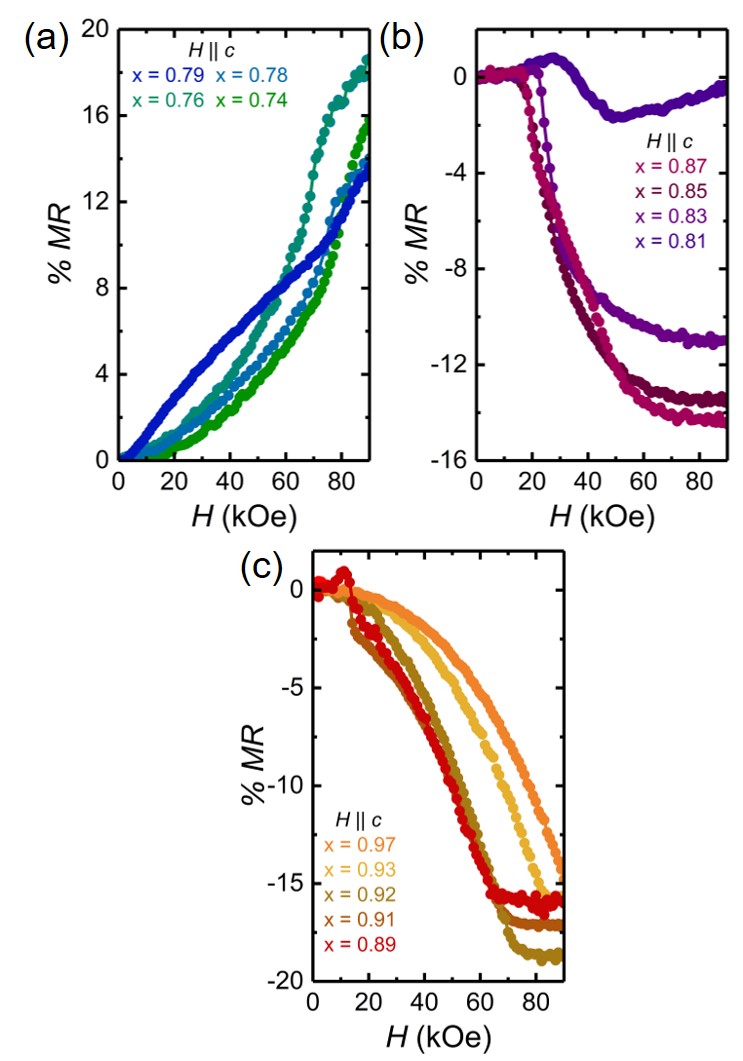}
    \caption[]{Transverse magnetoresistance of LaMn$_x$Sb$_2$ samples measured at 2 K with \textit{H} $\parallel$ \textit{c} and \textit{I} $\perp$ \textit{c}. The panels show data for different ranges of \textit{x} as follows: (a) \textit{x} = 0.74--0.79, (b) \textit{x} = 0.81--0.87, and (c) \textit{x} = 0.89--0.97.}
    \label{RH_2K}
\end{figure}

\subsubsection{Composition dependent magnetic phase diagram of LaMn$_x$Sb$_2$}

\begin{figure}[!t]
    \centering
    \includegraphics[width=\linewidth]{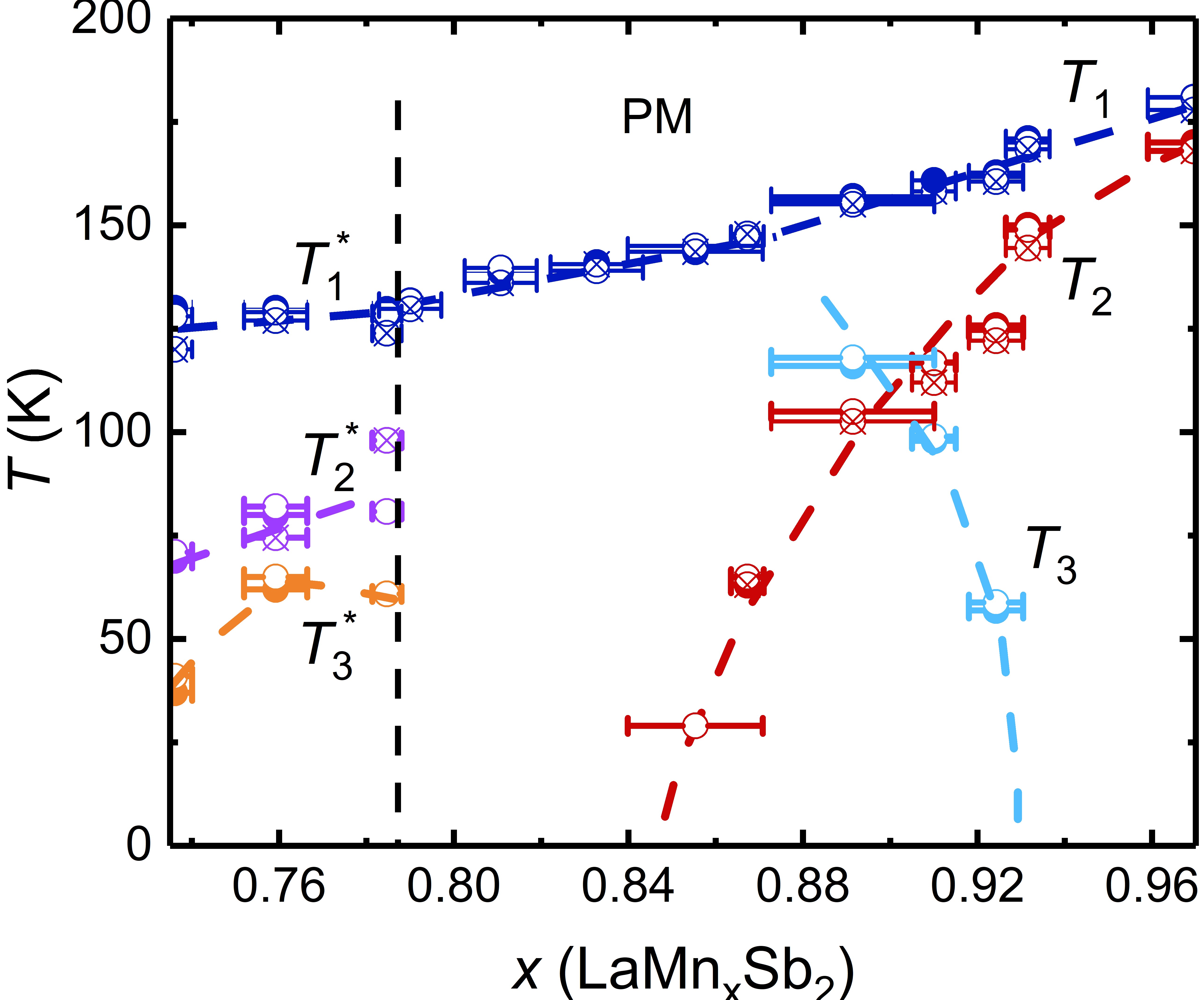}
    \caption[]{Temperature--composition (\textit{x}) phase diagram for LaMn$_x$Sb$_2$. The transition temperatures were determined by the local maxima in d($\chi$T)/d\textit{T} (filled points correspond to \textit{H} $\parallel$ \textit{c} and open points to \textit{H} $\perp$ \textit{c}) and d\textit{R}/d\textit{T} (crossed points). The vertical dashed black line on the left marks the structure change below \textit{x} $\leq$ 0.78, and the asterisk on the transition labels also refer to the different crystal structure.}
    \label{T-x}
\end{figure}

Figures \ref{MT_1kOe} and \ref{RT_all} respectively show the temperature dependent magnetization measured at \textit{H} = 1 kOe and the zero-field resistance of each LaMn$_x$Sb$_2$ sample. Likewise, Figures \ref{MH_5K} and \ref{RH_2K} show corresponding field dependent \textit{M}(\textit{H}) isotherms and the transverse magneoresistance resistance (\textit{MR}), defined as [\textit{R}(\textit{H})--\textit{R}(0)]/\textit{R}(0). The key results from Figures \ref{MT_1kOe}--\ref{RH_2K} are summarized in Figure \ref{T-x}, which shows a temperature-composition (\textit{T}--\textit{x}) phase diagram outlining the evolution of the various magnetic states observed for different compositions. Owing to the complexity of the data, we also show the data for each sample (individual plots for each value of \textit{x}) in Figures \ref{data_x74}--\ref{data_x97} in the Appendix. 

At a course level, the LaMn$_x$Sb$_2$ samples all have metallic transport behavior with relatively low residual resistance ratios, RRR = \textit{R}(300 K)/\textit{R}(2 K), consistent with the substantial vacancy concentrations. Both magnetic and transport results indicate that LaMn$_x$Sb$_2$ has rich, composition and temperature dependent magnetic behavior. The \textit{M/H} datasets in Figure \ref{MT_1kOe} show the onset of an initial antiferromagnetic feature at \textit{T}$_{1}$ = 130--180 K, where the transition temperatures increase with \textit{x}. Likewise, the resistance curves in Figure \ref{RT_all} all exhibit a clear loss of spin disorder scattering at \textit{T}$_1$, signalling the emergence of magnetic order. The values of \textit{T}$_1$ are consistent between magnetic and transport measurements, and below \textit{T}$_{1}$, most of the samples have additional transitions upon further cooling. In the following paragraphs, we outline the evolution of the magnetic states as the vacancy concentration is raised from \textit{x} = 0.74 to \textit{x} = 0.97. The discussion is organized such that we present data from Figures \ref{MT_1kOe}--\ref{RH_2K} for samples grouped into regions of \textit{x} that show similar behavior, roughly corresponding to a low-\textit{x} region with three transitions (\textit{x} = 0.74--0.78), a middle region with one or two transitions (\textit{x} = 0.79--0.87), and a high-\textit{x} region with two or three transitions (\textit{x} = 0.89--0.97). We outline this below in detail.

The magnetic data in Figures \ref{MT_1kOe}a and \ref{MT_1kOe}e and resistance data in Figure \ref{RT_all}a show that the low-\textit{x} LaMn$_x$Sb$_2$ samples with \textit{x} = 0.74--0.78 first enter a magnetically ordered state at \textit{T$_{1}$} $\approx$ 130 K, below which the behavior is complicated. For the \textit{H} $\parallel$ \textit{c} orientation (Figure \ref{MT_1kOe}a), \textit{M/H} reaches a peak at \textit{T}$_{1}$, and further cooling reveals a broad hump centered at $\approx$ 50--60 K, which may be indicative of a spin reorientation. For \textit{H} $\perp$ \textit{c} (Figure \ref{MT_1kOe}e), the susceptibility shows a kink at \textit{T$_{1}$} and monotonically increases upon further cooling, with two more kinks that appear at similar temperatures as the broad, hump-like feature observed in the \textit{H} $\parallel$ $\textit{c}$ data. Based on the d($\chi$\textit{T}/d\textit{T}) derivatives (Figures \ref{data_x74}b, \ref{data_x76}b, and \ref{data_x78}b in the appendix), the low \textit{x} samples each have two additional transitions below \textit{T}$_{1}$. Of these, only the higher temperature transition is observed in the resistance derivatives (Figure \ref{RT_all}d), which likely reflects the relatively low RRR. The lower temperature transitions do not trend monotonically with \textit{x}, and are both maximized at \textit{T}$_{2}$ $\approx$ 81 K and \textit{T}$_{3}$ $\approx$ 64 K for \textit{x} = 0.76. 

The magnetic and transport properties substantially change when \textit{x} $>$ 0.79. The most prominent difference is the abrupt disappearance of the two lower temperature transitions observed in the \textit{x} $\leq$ 0.78 samples. Likewise, the 2 K magnetoresistance found in Figure \ref{RH_2K} switches from positive to negative field dependence when \textit{x} $>$ 0.79. Finally, the \textit{RRR}s increase abruptly when \textit{x} $<$ 0.79 (see Figure \ref{RT_all}g). These sudden changes near \textit{x} = 0.78-0.79 are excellent agreement with the change from \textit{P}4/\textit{nmm} to \textit{I}$\bar{4}$2\textit{m} symmetry we observe when \textit{x} $\leq$ 0.78. The low-\textit{x} \textit{I}$\bar{4}$2\textit{m} structure, with three distinct Mn sites, apparently allows for two additional transitions below \textit{T}$_{1}$, which both abruptly vanish when the structure changes to the \textit{P}4/\textit{nmm} variant. Likewise, the partial ordering of the Mn vacancies in the \textit{I}$\bar{4}$2\textit{m} arrangement is consistent with the higher \textit{RRR}s observed for the lowest \textit{x} samples, despite nominally having more vacancies.

For LaMn$_x$Sb$_2$ samples with \textit{x} = 0.81--0.87, the temperature dependence of \textit{M/H} remains qualitatively similar to the lower \textit{x} samples, as \textit{M/H} reaches a peak at \textit{T}$_{1}$ when \textit{H} $\parallel$ \textit{c} and increases rapidly below the transition when \textit{H} $\perp$ \textit{c}. The samples with \textit{x} = 0.85 and \textit{x} = 0.87 undergo a second transition, \textit{T}$_2$, below \textit{T}$_1$ (see the arrows marking \textit{T}$_2$ in Figure \ref{MT_1kOe}f and in Figures \ref{RT_all}b and \ref{RT_all}e). \textit{T}$_{2}$ increases from 29 K to 65 K as \textit{x} is increased from 0.85 to 0.87. We note that separate magnetic and transport data was recently published by Yang et al. for samples with \textit{x} = 0.84 and \textit{x} = 0.86.\autocite{PhysRevB.107.115150} Our measurements are largely in agreement with the earlier data; however, Yang et al. observe only a single transition in both samples, whereas we clearly observe two transitions when \textit{x} $\geq$ 0.85. Given the otherwise good agreement between measurements, this discrepancy likely reflects the uncertainty in the composition, in which the samples reported by Yang et al. may have slightly lower values of \textit{x} than the corresponding samples reported here.

The behavior of \textit{M/H} again significantly changes when \textit{x} $\geq$ 0.89. First, Figures \ref{MT_1kOe} and \ref{MH_5K} both suggest that the easy direction switches from axial (moments oriented along \textit{c}) to planar (moments within the \textit{ab}-plane) when \textit{x} $>$ 0.89. This change is most clear in the \textit{M}(\textit{H}) isotherms shown in Figure \ref{MH_5K}. Here, samples with \textit{x} $\leq$ 0.89 undergo metamagnetic transitions when \textit{H} $\parallel$ \textit{c}, and the onset field moves to lower \textit{H} as \textit{x} increases. When \textit{x} $\geq$ 0.91, metamagnetism is instead observed when the field is applied in the \textit{ab}-plane (see the inset to Figure \ref{MH_5K}f), and the onset fields shift to higher \textit{H} as \textit{x} is increased. The change in sample orientation for which metamagnetism occurs clearly suggests the orientation of the ordered moment changes from axial to planar as \textit{x} $>$ 0.89. 

Furthermore, the behavior of \textit{M/H} at \textit{T}$_1$ changes when \textit{x} $\geq$ 0.89. Unlike the lower \textit{x} samples for which a peak in \textit{M/H} is observed at \textit{T}$_{1}$ when \textit{H} $\parallel$ \textit{c}, the \textit{M/H} for \textit{x} $\geq$ 0.89 samples have a more subtle feature at \textit{T}$_1$. Here, instead of a peak, the \textit{H} $\parallel$ \textit{c} and \textit{H} $\perp$ \textit{c} datasets begin to split away from each other (see Figures \ref{data_x89}--\ref{data_x97} in the appendix); however, \textit{M/H} continues to increase with further cooling in both sample orientations. When \textit{x} = 0.89--0.92, two lower temperature transitions are observed below \textit{T}$_1$, as denoted by the arrows in Figures \ref{MT_1kOe}d, \ref{MT_1kOe}h, and \ref{RT_all}f. In particular, Figure \ref{MT_1kOe}d shows that \textit{M/H} undergoes a very sharp drop at \textit{T}$_2$ for \textit{x} = 0.89 and at \textit{T}$_3$ for \textit{x} = 0.91 and 0.92. The step-like drop suggests a first order transition, and Figure \ref{T3_hysteresis} in the appendix shows that \textit{T}$_3$ has measurable hysteresis between warming/cooling sweeps, confirming that this transition is first-order 

Finally, the highest \textit{x} samples with \textit{x} $>$ 0.93 enter a magnetically ordered state at \textit{T}$_1$ = 170--180 K and undergo a second transition at \textit{T}$_2$ = 150--170 K where \textit{M/H} reaches a maximum in both sample orientations. Below the peak, \textit{M/H} falls monotonically when \textit{H} $\perp$ \textit{c} and has more complicated behavior in the \textit{H} $\parallel$ \textit{c} orientation, with a prominent minima near 140-150 K. These magnetization data suggest that samples with \textit{x} $\geq$ 0.93 undergo two closely spaced transitions below \textit{T}$_{1}$. However, field dependent measurements shown in Figure \ref{MT_RT_H_x93} in the appendix suggest the lowest temperature transition is a finite-field phase that is absent at lower fields but starts to become observable at $\approx$ 1 kOe and separates from the \textit{T} $\approx$ 150 K transition by moving to lower temperatures as the field is increased towards 10 kOe. Likewise, zero-field resistance measurements (see Figure \ref{RT_all}) show only two transitions. We therefore conclude that when \textit{x} $\geq$ 0.93, LaMn$_x$Sb$_2$ has two intrinsic zero-field transitions as denoted in the phase diagram, but shows rich field-dependent behavior where several new phases are observed at \textit{H} $>$ 1 kOe. We note that for samples with \textit{x} $<$ 0.93, measurements at both lower and higher fields are not susbstantially different than the \textit{H} = 1 kOe data presented in Figure \ref{MT_1kOe}.

\begin{figure*}[!t]
    \centering
    \includegraphics[width=\linewidth]{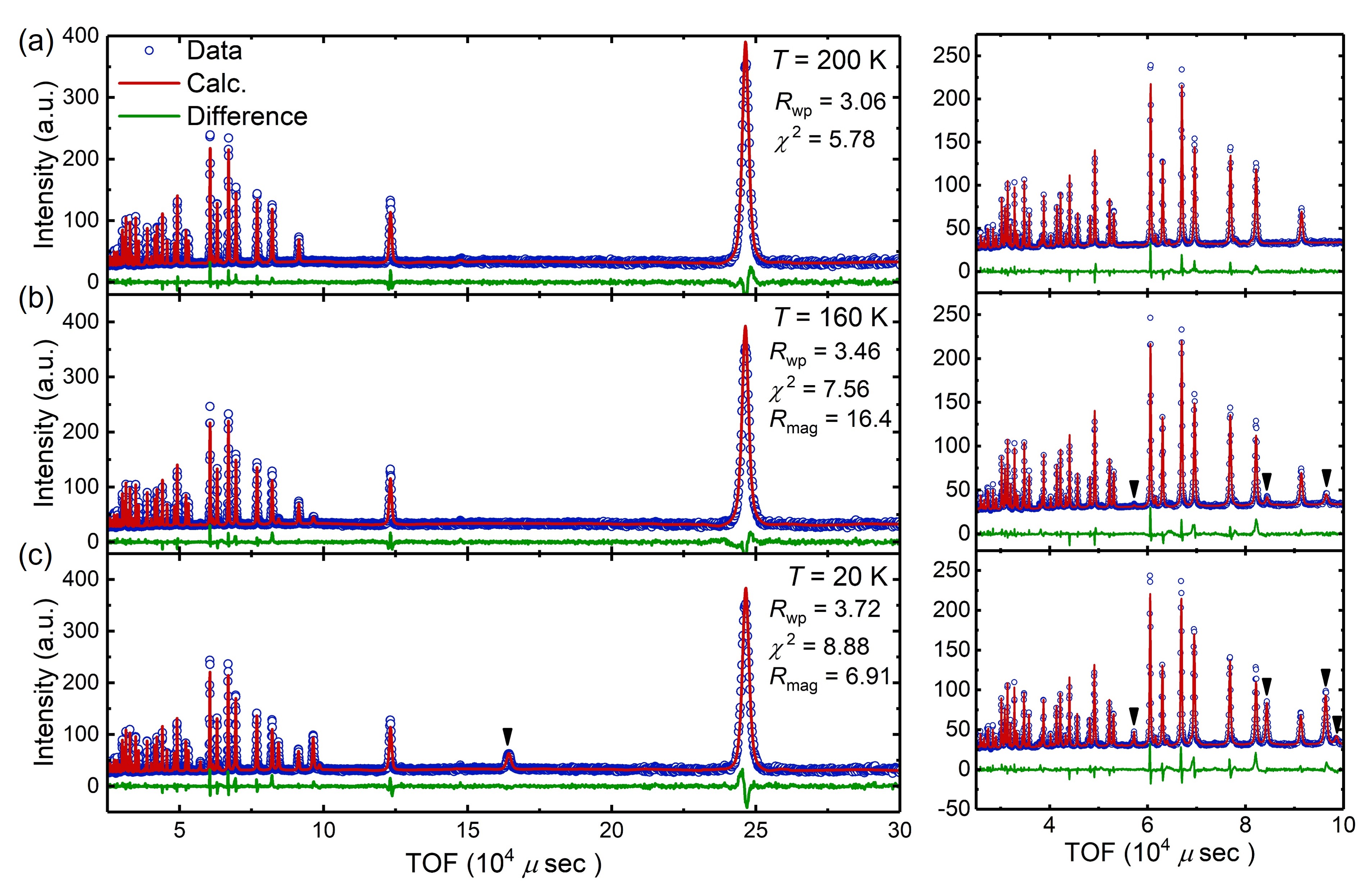}
    \caption[]{Time of flight (TOF) powder neutron diffraction patterns for \textit{x} = 0.93 LaMn$_x$Sb$_2$ collected at (a) 200 K, (b) 160 K, and (c) 20 K. The left panels are the full patterns and the right panels are a close-up view of the low time of flight (TOF) data. The black arrows mark the magnetic Bragg peaks not present in the 200 K data.}
    \label{neutron}
\end{figure*}

Figure \ref{T-x} summarizes the above results in a temperature composition (\textit{T}--\textit{x}) phase diagram. The phase diagram highlights the complexity of the magnetism in LaMn$_x$Sb$_2$, showing there are at least six magnetically ordered states, depending on the temperature and \textit{x}. The samples with closest to full Mn occupancy, \textit{x} = 0.93--0.97, undergo two magnetic transitions, where both transition temperatures \textit{T}$_1$ and \textit{T}$_2$ fall with decreasing \textit{x}. Starting near \textit{x} $\approx$ 0.92, there is third transition, which is first-order and occurs at higher temperatures as \textit{x} decreases, such that \textit{T}$_3$ moves above \textit{T}$_2$ at \textit{x} $\approx$ 0.9. 

The magnetic behavior distinctly changes at \textit{x} = 0.89. Based on the anisotropy observed in temperature and field dependent magnetization data, the direction of the ordered moment shifts from planar to axial orientation when \textit{x} $\leq$ 0.89. At intermediate \textit{x} = 0.79--0.87, \textit{T}$_1$ and \textit{T}$_2$ continue to fall monotonically as \textit{x} is lowered, and \textit{T}$_2$ is not observed below \textit{x} = 0.85. Further decreasing the Mn occupancy below \textit{x} $\leq$ 0.78 results in a change in the crystal structure from a \textit{P}4/\textit{nmm} arrangement with a crystallographically disordered Mn sublattice to a \textit{I}$\bar{4}$2\textit{m} structure with partial ordering of the Mn vacancies. The change in crystal structure is marked by the vertical dashed line in Figure \ref{T-x}. The structural transition results in the appearance of two low temperature transitions below \textit{T}$_1$ and a crossover from negative to positive magnetoresistance regimes at the lowest \textit{x}.


\noindent
\subsection{Magnetic Structures of LaMn$_x$Sb$_2$ for \textit{x} = 0.93}

Considering the possibility for electronic structure topology associated with the Sb net, the ability to tune between numerous magnetically ordered phases by varying the Mn occupancy \textit{x} may allow access to multiple topological states in LaMn$_x$Sb$_2$. Ultimately, neutron diffraction experiments covering each of the salient regions of the phase diagram are needed to provide the magnetic ordering wave vector and magnetic moment direction of each region that will lead to understanding of what topological phases can occur within the LaMn$_x$Sb$_2$ system. Such work is beyond the scope of the present text, so we limit ourselves here to the discusion of the highest \textit{x} region closest to the ideal LaMnSb$_2$ (\textit{x} = 1) compound.

Figure \ref{neutron} shows powder neutron diffraction patterns for a \textit{x} = 0.93 sample that were measured at 200 K, 160 K, and 20 K, corresponding to the paramagnetic state and two ordered states found in the highest \textit{x} region of the phase diagram. At 200 K, the neutron data shown in Figure \ref{neutron}a are consistent with a paramagnetic state, since all observed Bragg peaks correspond to the nuclear peaks anticipated for the \textit{P}4/\textit{nmm} structure. Several very weak reflections can also be indexed to MnSb, as was the case in the PXRD data. After cooling to 160 K, Figure \ref{neutron}b shows the emergence of three new peaks not found in the 200 K paramagnetic data, pointing to a reduction of symmetry associated with the onset of long-range antiferromagnetic order. Likewise, additional Bragg peaks are again observed in the 20 K pattern shown in Figure \ref{neutron}c. Because the 90 K (below both \textit{T}$_1$ and \textit{T}$_2$) synchrotron PXRD patterns discussed previously (Figure \ref{SynPXRD}c and \ref{SynPXRD}d) show no peaks beyond those expected from the \textit{P}4/\textit{nmm} structure, the new reflections observed in the neutron data must be associated with the magnetic order.

The magnetic reflections in the 160 K data (Figure \ref{neutron}b) are consistent with a propagation vector of $\tau$ = (0 0 1/2), corresponding to a magnetic unit cell that is double the chemical unit cell along the \textit{c}-direction. The additional Bragg peaks are (1 0 \textit{L}), where \textit{L} = (2\textit{n}+1)/2), and \textit{n} is an integer. The allowed magnetic structures were determined by group-subgroup analysis of the propagation vector $\tau$ = (0 0 1/2) applied to the high temperature crystal structure using \textit{SARAh}-Representational Analysis.\autocite{wills2000new} The symmetry analysis produces four allowed irreducible representations (IRs), and their associated basis vectors (BV) are given in Table \ref{BVtable}. 

Of the four IR's, $\Gamma_3$ and $\Gamma_9$ describe \textit{A}-type antiferromagnetic structures in which the moments are respectively aligned along the \textit{c}-axis ($\Gamma_3$) or within the \textit{ab}-plane ($\Gamma_9$). The magnetic structures corresponding to these IR's can be ruled out, as $\Gamma_3$ fails to capture all of the observed magnetic Bragg peaks, whereas $\Gamma_9$ leads to expected peaks that are not found in the 160 K diffraction pattern. The $\Gamma_6$ and $\Gamma_{10}$ IR's are both consistent with the observed Bragg peaks in Figure \ref{neutron}b and correspond to \textit{G}-type structures with axial and planar moment directions, respectively. Figure \ref{neutron}b shows the refinement using the planar \textit{G}-type structural model corresponding to $\Gamma_{10}$. The simulated pattern matches reasonably well with the experimental data, with \textit{R}$_{\text{wp}}$ = 3.46 and \textit{R}$_{\text{mag}}$ = 16.4, indicating a satisfactory refinement. Using the alternative, axial \textit{G}-type structure described by $\Gamma_6$ leads to a slightly worse refinement with \textit{R}$_{wp}$ = 3.60 and \textit{R}$_{\text{mag}}$ = 19.2. To fully resolve the ambiguity, we also conducted single crystal neutron diffraction, and the data is shown in Figure \ref{SCneutron} in the appendix. The single crystal neutron diffraction measurements indicate the moment direction is within the \textit{ab}-plane, ruling out the $\Gamma_6$ structure.

\begin{figure}[!t]
    \centering
    \includegraphics[width=\linewidth]{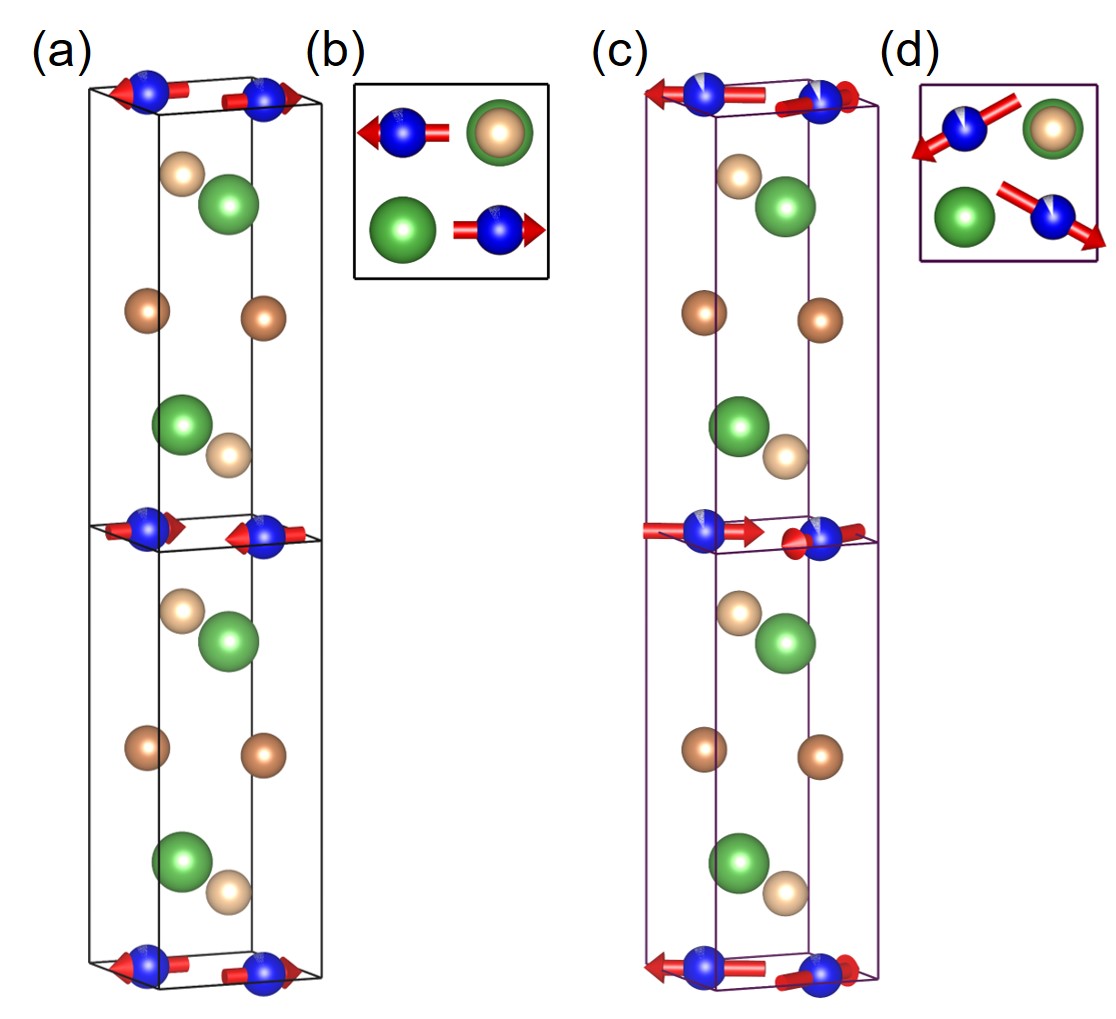}
    \caption[]{Magnetic structure for \textit{x} = 0.93 LaMn$_x$Sb$_2$ at (a) and (b) 160 K, (c) and (d) 20 K. (b) and (d) are perspectives looking down the \textit{c}-axis showing the in-plane rotation of the Mn moments in the 20 K structure. The color code is as follows: green = La, blue = Mn, and brass = Sb. The red arrows depict the orientation of the Mn moments.}
    \label{MagStructure}
\end{figure}

The refined magnetic structure is illustrated in Figures \ref{MagStructure}a and \ref{MagStructure}b. At 160 K, LaMn$_{0.93}$Sb$_2$ adopts a G-type antiferromagnetic structure in which the Mn moments align antiparallel within the basal-plane and stack in alternating directions along the \textit{c}-axis. As noted above, this structure is described by the $\Gamma_{\text{10}}$ IR, and the refined moment is $\mu_{\text{tot}}$ = 1.4(8) $\mu_{\text{B}}$/Mn. Very recently, Yang et al. reported neutron diffraction data for a LaMn$_x$Sb$_2$ sample with \textit{x} = 0.86, and found that this composition also adopts a \textit{G}-type antiferromagnetic structure, but with spins oriented along the \textit{c}-axis.\autocite{PhysRevB.107.115150} These results are consistent with our magnetic measurements that indicate the direction of the ordered moment changes from planar to axial alignment when \textit{x} $\leq$ 0.89. 

The neutron powder pattern collected at 20 K (Figure \ref{neutron}c) shows several new Bragg peaks, marked by arrows in Figure \ref{neutron}c, indicating the magnetic structure is different than that observed at 160 K. The magnetic Bragg peaks in the 20 K neutron data can again be indexed to a propagation vector of $\tau$ = (0 0 1/2); however, both $\Gamma_9$ and $\Gamma_{10}$ are needed to capture all of the observed reflections. Refinement using the basis vectors $\psi_4$ and $\psi_5$, or equivalently, $\psi_3$ and $\psi_6$ of $\Gamma_9$ and $\Gamma_{10}$ respectively produces the magnetic structure shown in Figure \ref{MagStructure}c, with reasonable statistics of \textit{R}$_{\text{wp}}$ = 3.72 and \textit{R}$_{\text{mag}}$ = 6.91.

Upon cooling below \textit{T}$_2$, we find that the \textit{G}-type magnetic structure distorts by a rotation of the magnetic moments within the basal plane as seen by comparing Figures \ref{MagStructure}b and \ref{MagStructure}d. The direction of the rotation alternates between Mn layers such that the moments remain antiparallel when stacked along the \textit{c}-axis. The distortion corresponding to rotation of the moment directions within the plane requires the addition of the $\Gamma_9$ IR to the collinear structure described by $\Gamma_{10}$ that is observed at 160 K. This structure is consistent with the experimental magnetic anisotropy (see Figures \ref{MT_1kOe} or \ref{data_x93}), which also indicated that the moment direction is primarily within the basal plane. The refinement suggests the total moment associated with the Mn atoms is 3.51(2) $\mu_{\text{B}}$ at 20 K. Given our 2 K \textit{M}(\textit{H}) measurements shown in Figure \ref{MH_5K} indicate a maximum moment of 1.4 $\mu_{\text{B}}$/Mn at 70 kOe, these results likely suggest additional metamagnetism should occur at higher fields.

\begin{table}[!t]
\caption{Irreducible representations and corresponding basis vectors (BV) for the space group \textit{P}4/\textit{nmm} with $\tau$ = (0,~ 0,~ 1/2). The decomposition of the magnetic representation for the Mn site (3/4,~ 1/4,~ 0) is $\Gamma_{3}^{1} + \Gamma_{6}^{1} + \Gamma_{9}^{2} + \Gamma_{10}^{2}$. The atoms of the nonprimitive basis are chosen to be 1: (3/4,~ 1/2,~ 0), 2: (1/2,~ 3/4,~ 0).}
\resizebox{\linewidth}{!}{\begin{tabular}{ccc|cccccc}
\toprule
  IR  &  BV  &  Atom & \multicolumn{6}{c}{BV components}\\
      &      &             &$m_{\|a}$ & $m_{\|b}$ & $m_{\|c}$ &$im_{\|a}$ & $im_{\|b}$ & $im_{\|c}$ \\
\hline
$\Gamma_{3}$ & $\psi_{1}$ &      1 &      0 &      0 &      8 &      0 &      0 &      0  \\
             &              &      2 &      0 &      0 &      8 &      0 &      0 &      0  \\
$\Gamma_{6}$ & $\psi_{2}$ &      1 &      0 &      0 &      8 &      0 &      0 &      0  \\
             &              &      2 &      0 &      0 &     -8 &      0 &      0 &      0  \\
$\Gamma_{9}$ & $\psi_{3}$ &      1 &      4 &      0 &      0 &      0 &      0 &      0  \\
             &              &      2 &      4 &      0 &      0 &      0 &      0 &      0  \\
             & $\psi_{4}$ &      1 &      0 &     -4 &      0 &      0 &      0 &      0  \\
             &              &      2 &      0 &     -4 &      0 &      0 &      0 &      0  \\
$\Gamma_{10}$ & $\psi_{5}$ &      1 &      4 &      0 &      0 &      0 &      0 &      0  \\
             &              &      2 &     -4 &      0 &      0 &      0 &      0 &      0  \\
             & $\psi_{6}$ &      1 &      0 &      4 &      0 &      0 &      0 &      0  \\
             &              &      2 &      0 &     -4 &      0 &      0 &      0 &      0  \\
             \bottomrule
\end{tabular}}%
\label{BVtable}
\end{table}

\noindent
\subsection{Electronic Band Structure}

\begin{figure}[!t]
    \centering
    \includegraphics[width=\linewidth]{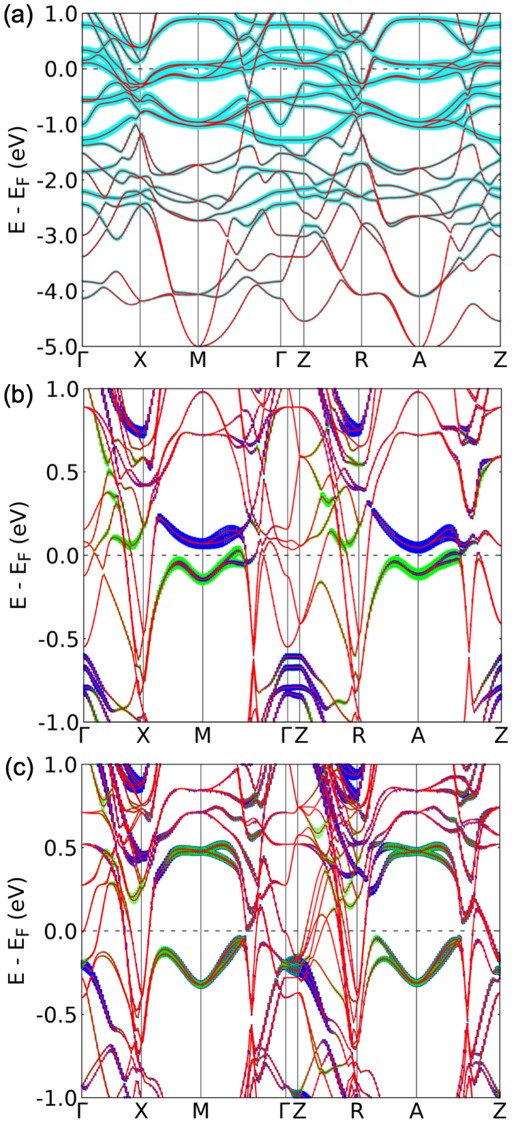}
    \caption[]{Electronic band structure for LaMnSb$_2$ (assuming \textit{x} = 1) calculated in (a) the paramagnetic state, (b) the collinear \textit{G}-type antiferromagnetic structure, and (c) the non-collinear antiferromagnetic structure. The cyan shading in (a) represents the projection of the Mn-3\textit{d} orbitals. The blue and green shading in (b) and (c) represent the Mn \textit{d}$_{xz}$ and \textit{d}$_{yz}$ orbital projections, respectively.}
    \label{BandStructure}
\end{figure}

Having determined the magnetic structures for the high-\textit{x} region of the phase diagram, we used density functional theory (DFT) to calculate the electronic band structure of LaMn$_x$Sb$_2$ in the paramagnetic and magnetic states, assuming \textit{x} = 1 (no Mn vacancies). Figure \ref{BandStructure}a shows the band structure calculated with spin-orbit coupling in the paramagnetic (or nonmagnetic) phase. The projection overlaid on the band structure shows that the states around the Fermi level (\textit{E}$_{\text{F}}$) are mostly derived from the Mn 3\textit{d} orbitals, which produce a small band dispersion and large density states near the \textit{E}$_{\text{F}}$ to induce magnetic exchange splitting. In contrast, the states with large band dispersion are derived from Sb 5\textit{p} orbitals and well below the \textit{E}$_{\text{F}}$, except for along certain directions, such as the \textit{M}-$\Gamma$ and \textit{A}–-\textit{Z} directions, where they also overlap with Mn 3d bands.

Figures \ref{BandStructure}b and \ref{BandStructure}c display the band structures zoomed in near the \textit{E}$_{\text{F}}$ for the collinear \textit{G}-type antiferromagnetic structure and lower temperature non-collinear antiferromagnetic state, respectively. The DFT calculations indicate that the non-collinear structure is 28 meV/f.u. more stable than the collinear structure, consistent with the experimental finding of the non-collinear ground state. Likewise, the calculated magnetic moment is 3.8 $\mu_{\text{B}}$/Mn, in reasonably agreement with the 3.5 $\mu_{\text{B}}$ moment inferred from the 20 K neutron data. The valence bands just below the \textit{E}$_{\text{F}}$ near the \textit{M} and \textit{A} points are derived from the Mn \textit{d}$_{xz}$ orbitals (green), while the conduction band just above the \textit{E}$_{\text{F}}$ are derived from the Mn \textit{d}$_{yz}$ orbitals (blue). This splitting reflects that the magnetic moments in the collinear state are along the \textit{a}-axis. Entering the lower temperature non-collinear state, as shown in Figure \ref{BandStructure}c, the valence and conduction bands near the \textit{M} and \textit{A} points now both have a mixture of \textit{d}$_{xz}$ and \textit{d}$_{yz}$ character, which favors a larger splitting and gives a more stable magnetic state than the collinear case. Such splitting of $\approx$ 1 eV around $E_{\text{F}}$ among the different Mn 3\textit{d} orbitals is smaller than the exchange splitting of $\approx$ 3 eV for the spin moment of Mn for each orbital. For the collinear configuration, the spin moment is also near 3.8 $\mu_{\text{B}}$/Mn. The smaller measured apparent moment at \textit{T} = 160 K is due to partially random orientation of moments at elevated temperature. Because our magnetization isotherms obtained at \textit{T} = 2 K only reach maximum values of $\approx$ 1.4 $\mu_{\text{B}}$/Mn, the larger moment inferred by both neutron diffraction and the DFT calculations strongly suggest that additional metamagnetic transitions will occur above 70 kOe.

The calculated band structures in Figure \ref{BandStructure}b and \ref{BandStructure}c also show Dirac-like band dispersion from the gaped nodal lines along the \textit{M}–$\Gamma$ and \textit{A}–\textit{Z} directions located $\approx$ 0.6 eV below the \textit{E}$_{\text{F}}$. These Dirac-like bands are derived from the Sb 5\textit{p} orbitals that make up the square net and are reminiscent of the bands near \textit{E}$_{\text{F}}$ in the isostructural CaMnBi$_2$. In LaMn$_x$Sb$_2$, the presence of trivalent La instead of divalent Ca should electron dope the material and push \textit{E}$_{\text{F}}$ to higher energy in LaMn$_x$Sb$_2$ (assuming a rigid band model), which is consistent with our calculated results showing \textit{E}$_{\text{F}}$ well above the Dirac-like dispersion. In both magnetic phases, the calculations show gaped Dirac-like bands $\approx$ 0.6 eV below the Fermi level. This finding unfortunately suggests the LaMn$_x$Sb$_2$ samples with \textit{x} close to 1 are unlikely to display the topological physics associated with the Sb \textit{p}-orbital derived Dirac-like bands on the square lattice. Because each Mn vacancy will in principle introduce two holes into the valence band, \textit{E}$_{\text{F}}$ should move closer to the Dirac-like dispersion in the samples with lower \textit{x}.

\section{Discussion}

Our work shows that it is possible to chemically manipulate the Mn occupacy in LaMn$_x$Sb$_2$. The high vacancy concentrations endemic to LaMn$_x$Sb$_2$ likely arise from an attempt to achieve charge balance, which can be seen by comparison with other \textit{A}\textit{TM}Sb$_2$ compounds. In LaMn$_x$Sb$_2$, there are two unique Sb atoms in each unit cell, one bonded to the Mn atoms in tetrahedrally coordinated layers, and the second bonded only to other Sb atoms and forming a 2D square net (see Figure \ref{Structure}). According to the electron counting rules described by Hoffman,\autocite{a2000hypervalent} the formal charge on the antimony atoms in the Mn-Sb layers should be 3- and for those in the square net should be 1-, giving a net 4- charge from the anions. When both the transition metal and \textit{A} cations are divalent, for example \textit{TM} = Mn, Fe, Co, Ni, Zn and \textit{A} = Ca, Sr, Ba, Eu, Yb, the net positive charge balances the negative charge from the Sb, and vacancies are not reported in these \textit{ATMPn}$_2$ compounds,\autocite{PhysRevB.99.054435,liu2016nearly,li2016electron,kealhofer2018observation,masuda2016quantum,yi2017large}. Likewise, charge balance is also achieved in \textit{R}AgSb$_2$ and \textit{R}AuSb$_2$, where \textit{R} is a trivalent rare earth and \textit{TM} is the monovalent Ag or Au, and no (or considerably fewer, \textit{x} = 0.9--1 for Au) Ag/Au vacancies are reported.\autocite{myers1999systematic,wollesen1996ternary} On the other hand, the ATM$_{1-x}$Pn$_2$ combinations in which \textit{A} is trivalent and \textit{TM} is divalent are nominally electron rich for \textit{x} = 0. In these cases, the vacancies, with \textit{x} reported as low as 0.5,\autocite{leithe1994crystal,sologub1994crystal,wollesen1996ternary} move the total electron count closer to charge balance.

Here, we are able to grow single crystalline LaMn$_x$Sb$_2$ with \textit{x} spanning $\approx$ 0.74--0.97 simply by varying the composition of the starting melt. Whereas the distance between the Dirac bands and the Fermi level indicate LaMn$_x$Sb$_2$ is unlikely to display topological physics, our measurements demonstrate that LaMn$_x$Sb$_2$ is nonetheless a remarkably tuneable material in which the magnetic state is strongly dependent on the Mn occupancy. One possible explanation for this behavior is found in the electronic band structure. Figure \ref{BandStructure} shows a very flat band near \textit{E}$_F$ along \textit{X}--\textit{M}--\textit{G} which is derived from Mn \textit{d}-orbitals. Such a flat band should produce a spike in the density of states. In intermetallic compounds, the exchange interaction and proclivity towards magnetic order depend intimately on the density of states at the Fermi level DOS(\textit{E}$_F$); consequentially, increasing the number of Mn vacancies will lower \textit{E}$_F$, change DOS(\textit{E}$_F$), and likely alter the magnetic structure.
 
Another observation from our data is that compared to other \textit{A}Mn\textit{Pn}$_2$ materials, the magnetic ordering temperatures for LaMn$_x$Sb$_2$ are surprisingly low. Depending on the specific compound, the Mn sublattices in other \textit{A}Mn\textit{Pn}$_2$ compounds, where \textit{A} = Ca, Sr, Ba, Eu, or Yb, are all reported to order antiferromagnetically at \textit{T}$_N$ $\approx$ 270--350 K,\autocite{he2012giant,wang2012two,liu2017magnetic,liu2016nearly,park2011anisotropic,li2016electron,kealhofer2018observation,wang2018quantum,liu2017unusual,zhu2020magnetic,soh2019magnetic,yi2017large,PhysRevB.100.014437} temperatures approximately double the initial transitions (\textit{T}$_{1}$ = 130--180 K) observed here in LaMn$_x$Sb$_2$. The possibility of magnetic transitions above 300 K can likely be ruled out in LaMn$_x$Sb$_2$ by the neutron diffraction, as the patterns collected at 200 K show no additional reflections beyond the anticipated nuclear peaks, indicating there is no long range order above \textit{T}$_{1}$. Likewise, the temperature and field dependent magnetic data collected above \textit{T}$_1$ are consistent with paramagnetic behavior.

Whereas it may be tempting to relate the low ordering temperatures with the disorder associated with the Mn vacancies, this notion is inconsistent with the weak trend of \textit{T}$_{1}$ with \textit{x}, which suggests an extrapolated ordering temperature still below 200 K for a hypothetical disorder-free LaMnSb$_2$ (\textit{x} = 1). More realistically, the higher position of the Fermi level in LaMn$_x$Sb$_2$ compared to the other \textit{A}Mn\textit{Pn}$_2$ materials substantially changes the DOS(\textit{E}$_F$) and likely results in a relatively low ordering temperature.

\section{Summary and Conclusions}

We studied evolution of the magnetic and transport properties of LaMn$_x$Sb$_2$ as a function of Mn vacancy concentration \textit{x}. By altering the composition of ternary La--Mn--Sb melts, we are able to control the Mn occupancy and produce single crystals of LaMn$_x$Sb$_2$ with \textit{x} $\approx$ 0.74--0.97. Whereas some previous publications indicate LaMn$_x$Sb$_2$ may order ferromagnetically, our work unambiguously demonstrates that LaMn$_x$Sb$_2$ is a complex antiferromagnet. We provide evidence suggesting earlier reports of ferromagnetism can be explained by the presence of a small of MnSb impurity, which masks the intrinsic properties of LaMn$_x$Sb$_2$ unless it is carefully removed. Our transport and magnetic measurements show that LaMn$_x$Sb$_2$ has an rich composition-temperature phase diagram, first entering into an antiferomagnetic state at \textit{T}$_{1}$ $\approx$ 130--180 K and with six possible magnetic phases accessible by changing the temperature and Mn occupancy. Powder neutron diffraction data show that at high values of \textit{x} $\geq$ 0.93, LaMn$_x$Sb$_2$ first orders in a collinear \textit{G}-type antiferromagnetic arrangement, with moments aligned within the basal-plane, and this is followed by a second transition into a non-collinear state where the moments are rotated whithin the plane. Furthermore, we observe a change in the crystal structure from the high-\textit{x} \textit{P}4/\textit{nmm} arrangement to a low-\textit{x} \textit{I}$\bar{4}$2\textit{m} structure when \textit{x} $\leq$ 0.78. The change in structure is associated with partial ordering of the Mn vacancies and results in the sudden appearance of two new magnetic states below \textit{T}$_{1}$ and a crossover from negative to positive magnetoresistance regimes when \textit{x} falls below 0.79. Ultimately, additional neutron diffraction measurements will be needed to assess the numerous magnetic phases; however, given the ability to tune between six magnetically ordered states and two crystal structures, LaMn$_x$Sb$_2$ appears to be a rich playground for studying different phases within a single material system.

\vskip 0.25cm
\noindent
\textbf{\textit{Acknowledgements}}

Work at the Ames National Laboratory (TJS, AS, JMW, LLW, JS, SLB, and PCC) was supported by the U.S. Department of Energy, Office of Science, Basic Energy Sciences, Materials Sciences and Engineering Division. Ames National Laboratory is operated for the U.S. Department of Energy by Iowa State University under Contract No. DE-AC02-07CH11358. TJS, LLW, and PCC were supported by the Center for Advancement of Topological Semimetals (CATS), an Energy Frontier Research Center funded by the U.S. Department of Energy Office of Science, Office of Basic Energy Sciences, through Ames National Laboratory under its Contract No. DE-AC02-07CH11358 with Iowa State University. Use of the Advanced Photon Source at Argonne National Laboratory was supported by the U. S. Department of Energy, Office of Science, Office of Basic Energy Sciences, under Contract No. DE-AC02-06CH11357. A portion of this research used resources at the Spallation Neutron Source, a DOE Office of Science User Facility operated by the Oak Ridge National Laboratory.  The authors thank Tom Lograsso and Matt Kramer for useful discussions.

\vskip 0.25cm
\noindent
\textbf{\textit{Conflicts of Interest}}

The authors have no conflicts of interest to declare.

\printbibliography

\section{Appendix}

\noindent
\subsection{Removal of the MnSb second phase}

\begin{figure}[!t]
    \centering
    \includegraphics[width=\linewidth]{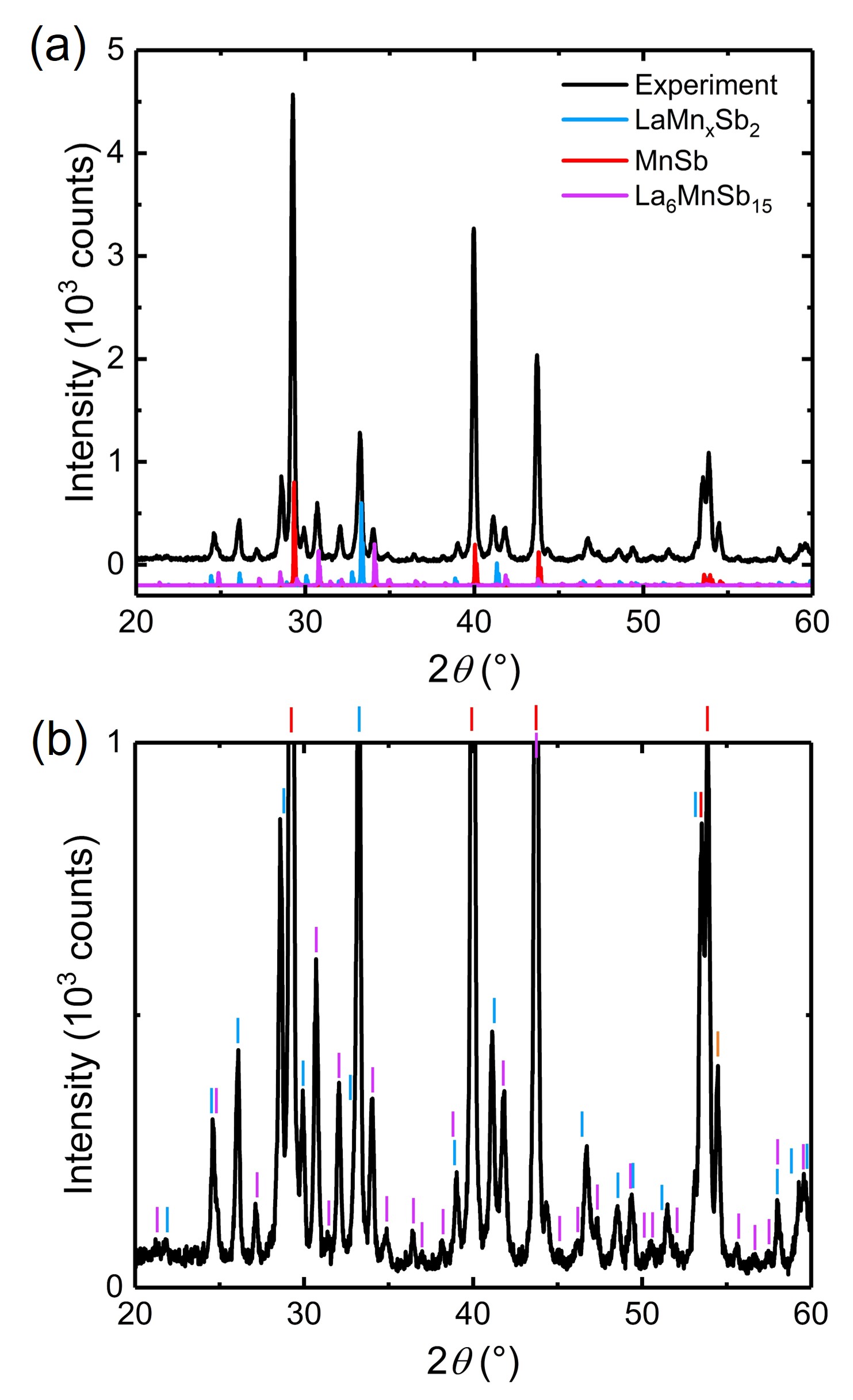}
    \caption[]{(a) Powder X-ray diffraction pattern obtained from the solidified decanted liquid that passed through the frit-disc and was captured by the catch crucible. (b) is a closeup of the low-intensity data to better see the weaker peaks. The experimental pattern is compared with the simulated diffraction patterns for LaMn$_x$Sb$_2$, MnSb, and La$_6$MnSb$_{15}$, showing the decant to be primarily composed of MnSb, with smaller quantities of LaMn$_x$Sb$_2$ and La$_6$MnSb$_{15}$. The vertical lines in (b) show the expected position of the Bragg peaks for each phase.}
    \label{SpinPXRD}
\end{figure}

As discussed in the crystal growth section, the LaMn$_x$Sb$_2$ crystals always formed with a small fraction of a MnSb impurity (see the right inset to Figure \ref{pxrd}) that is normally visible as small droplets or patches of residual self-flux on the surfaces of the crystals. Figure \ref{SpinPXRD} shows a powder X-ray diffraction pattern obtained from the solidified decant. The pattern indicates that the primary phase in the decanted liquid is MnSb, with smaller amounts of LaMn$_x$Sb$_2$ and the known ternary compound La$_6$MnSb$_{15}$. The high fraction of MnSb in the decant is consistent with our discussion in the main text and implies that the MnSb secondary phase found on our LaMn$_x$Sb$_2$ crystals forms from residual flux that is not completely removed during the centrifugation/decanting step. 

Because MnSb is a known ferromagnet with \textit{T}$_C$ $>$ 300 K,\autocite{okita1968crystal,teramoto1968existence} it was essential to eliminate the impurity to avoid obscuring the intrinsic features in the magnetic data. This is especially pertinent given that some of the limited data available also suggests that LaMn$_x$Sb$_2$ is ferromagnetic with a \textit{T}$_C$ $\approx$ 310 K.\autocite{sologub1995ternary} To remove the MnSb, we carefully polished each surface of our sample. After this cleaning step, the crystals were no longer attracted to a Nd$_2$Fe$_{14}$B magnet, suggesting LaMn$_x$Sb$_2$ is not intrinsically ferromagnetic at room temperature. 

\begin{figure}[!t]
    \centering
    \includegraphics[width=\linewidth]{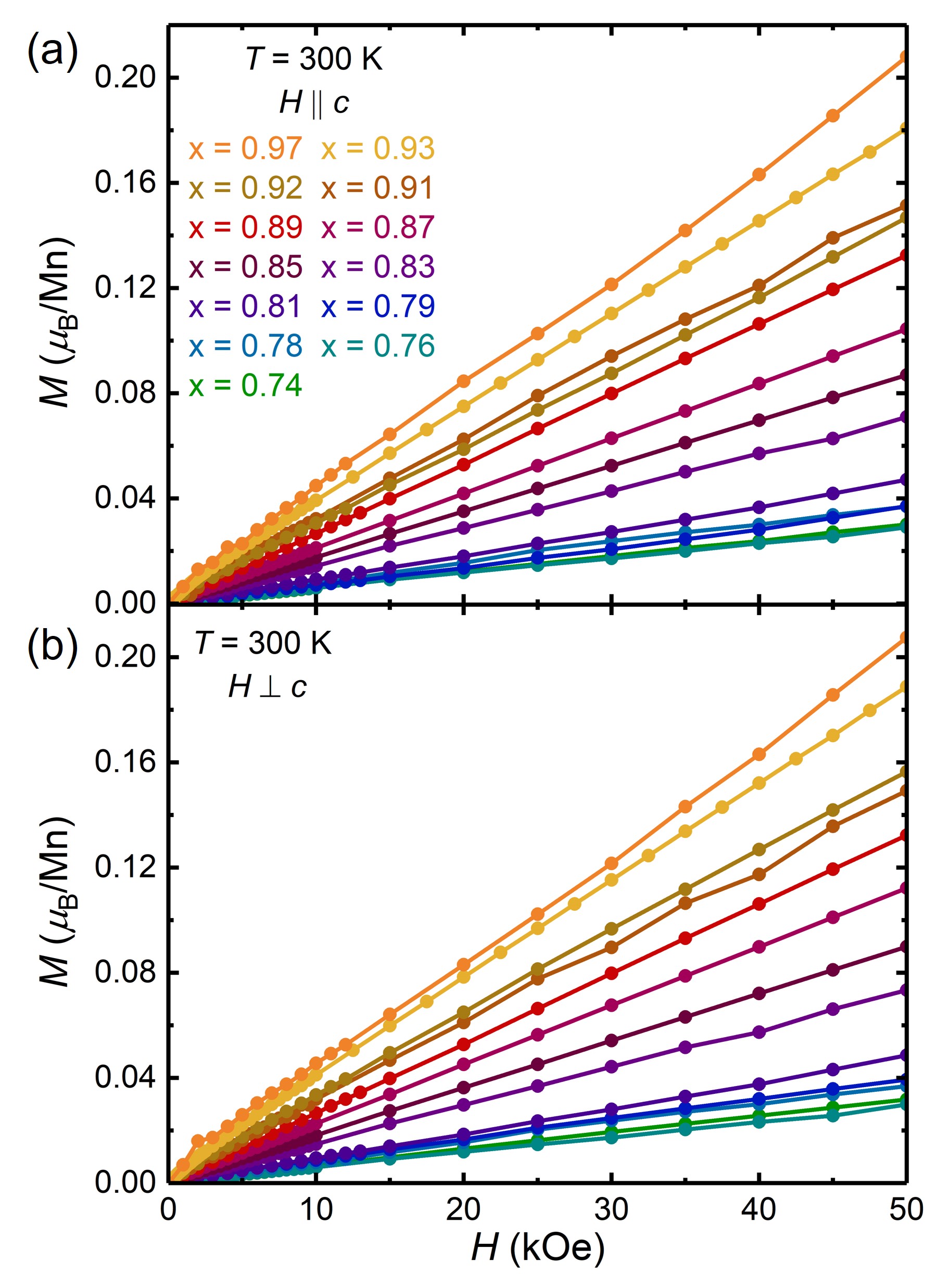}
    \caption[]{Field dependent Magnetization for LaMn$_x$Sb$_2$ measured at 300 K. (a) Results for samples oriented with \textit{H} $\parallel$ \textit{c} and (b) for \textit{H} $\perp$ \textit{c}.}
    \label{MH_300K}
\end{figure}

\begin{figure}[!t]
    \centering
    \includegraphics[width=\linewidth]{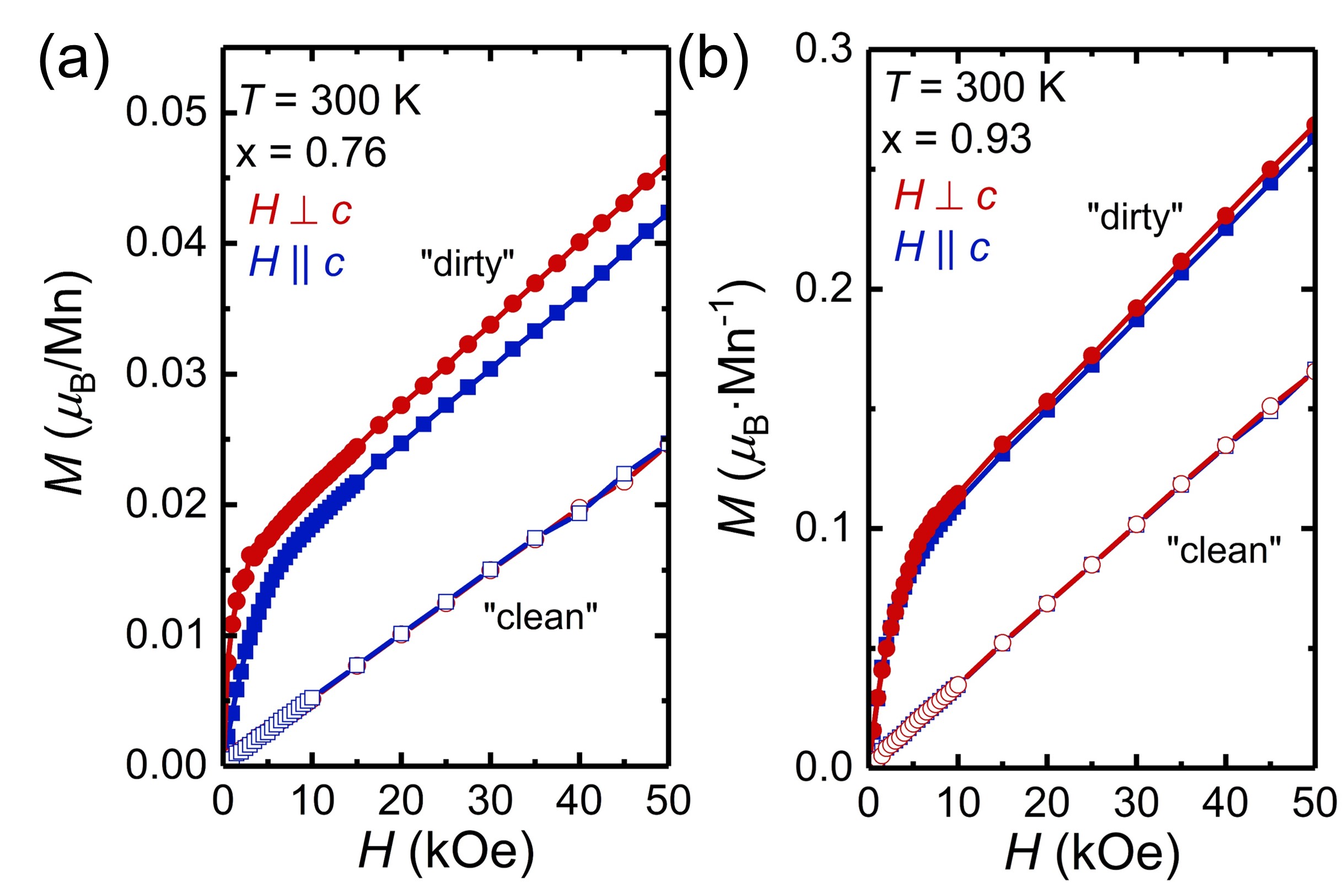}
    \caption[]{Field dependent magnetization data measured at 300 K on "dirty" samples of LaMn$_x$Sb$_2$ containing a MnSb impurity (solid points). The open points show the data after polishing all surfaces of the sample to remove the MnSb. (a) shows the data for x = 0.93 and (b) for x = 0.76. Both samples show clear signatures of ferromagnetic behavior at low fields, superimposed on the intrinsic linear behavior of the LaMn$_x$Sb$_2$.}
    \label{MH_dirty}
\end{figure}

Figures \ref{MH_300K}a and \ref{MH_300K}b show the field dependent magnetization \textit{M}(\textit{H}) curves collected on the cleaned LaMn$_x$Sb$_2$ at 300 K. The magnetization increases linearly with \textit{H} up to the highest fields, inconsistent with ferromagnetic order at room temperature. These results should be contrasted with the data found in Figure \ref{MH_dirty}, which shows \textit{T} = 300 K \textit{M}(\textit{H}) isotherms obtained on as-grown samples still containing residual MnSb on their surfaces. The magnetization of the MnSb contaminated samples increases rapidly upon application of a small field, and at $\approx$ 2 kOe, begins to show a more gradual, linear, field dependence that persists up to 55 kOe. The open points in Figure \ref{MH_dirty} show the magnetization of the same samples after polishing, demonstrating that the low field saturation is eliminated once the surfaces have been cleaned. Therefore, the results in Figure \ref{MH_dirty}a and \ref{MH_dirty}b can be interpreted as the signal from the ferromagnetic impurity (MnSb) imposed on the intrinsic antiferromagnetic properties of LaMn$_x$Sb$_2$. In both cases, the MnSb moments are quickly polarized by $\approx$2 kOe, which gives the steep rise in \textit{M} at low fields. Above 2 kOe, the linear increase of \textit{M} is the intrinsic signal from LaMn$_x$Sb$_2$ in the paramagnetic regime. 

Based on the above results, we conclude that LaMn$_x$Sb$_2$ is not a \textit{T}$_C$ $>$ 300 K ferromagnet. Considering that MnSb appears to readily form from La--Mn--Sb melts and is present as a second phase in all of our as-grown samples, it is very likely that the polycrystalline samples previously characterized also contained a small amount of MnSb that lead to the misattribution of ferromagnetism to LaMn$_x$Sb$_2$.

\begin{figure}[!t]
    \centering
    \includegraphics[width=\linewidth]{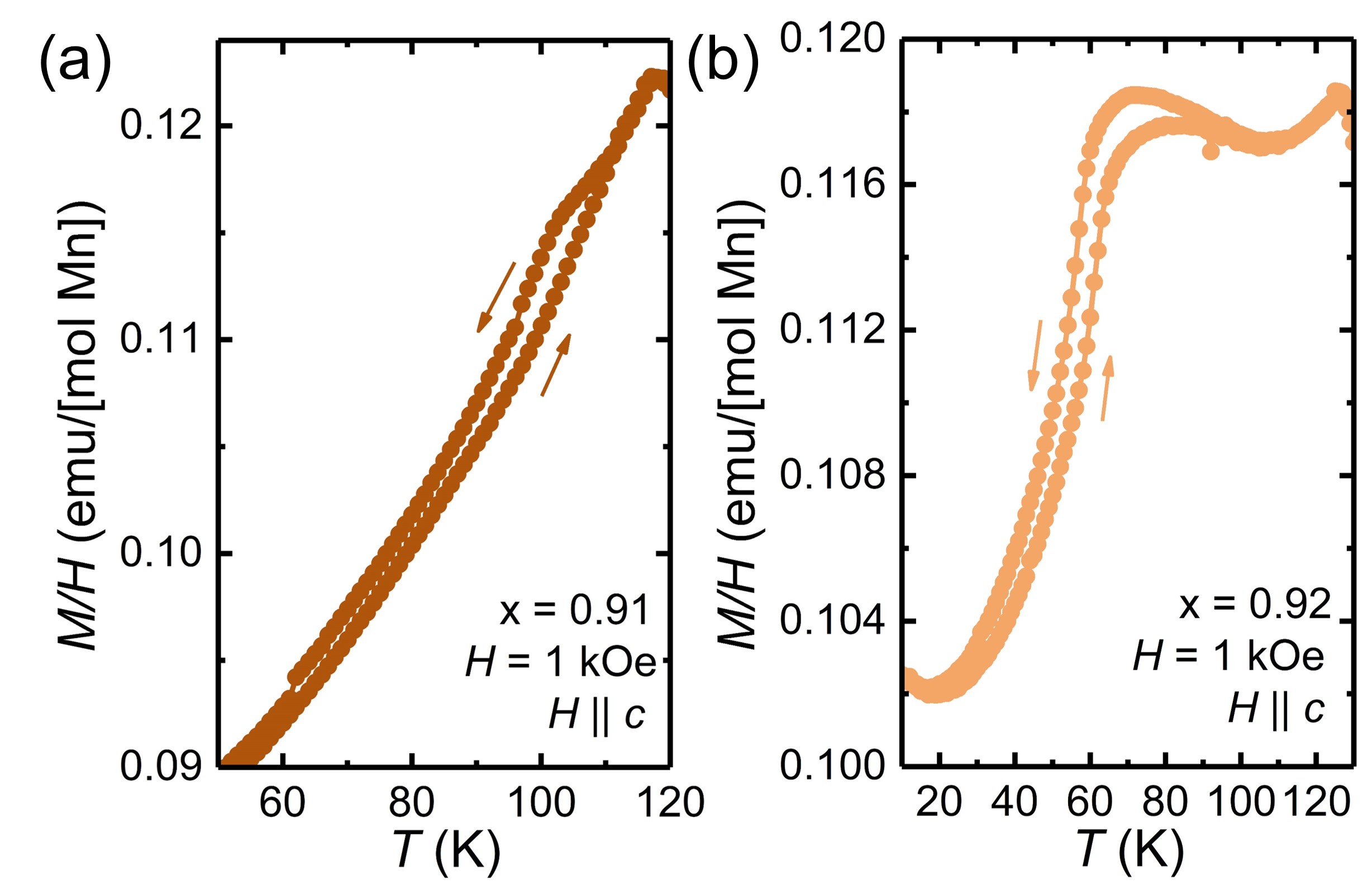}
    \caption[]{Close up view of \textit{M/H} around \textit{T}$_3$ in (a) \textit{x} = 0.91 and (b) \textit{x} = 0.92 LaMn$_x$Sb$_2$. Both datasets show measurable hysteresis between warming/cooling sweeps, indicating \textit{T}$_3$ corresponds to a first-order transition.}
    \label{T3_hysteresis}
\end{figure}

\begin{figure*}[!t]
    \centering
    \includegraphics[width=\linewidth]{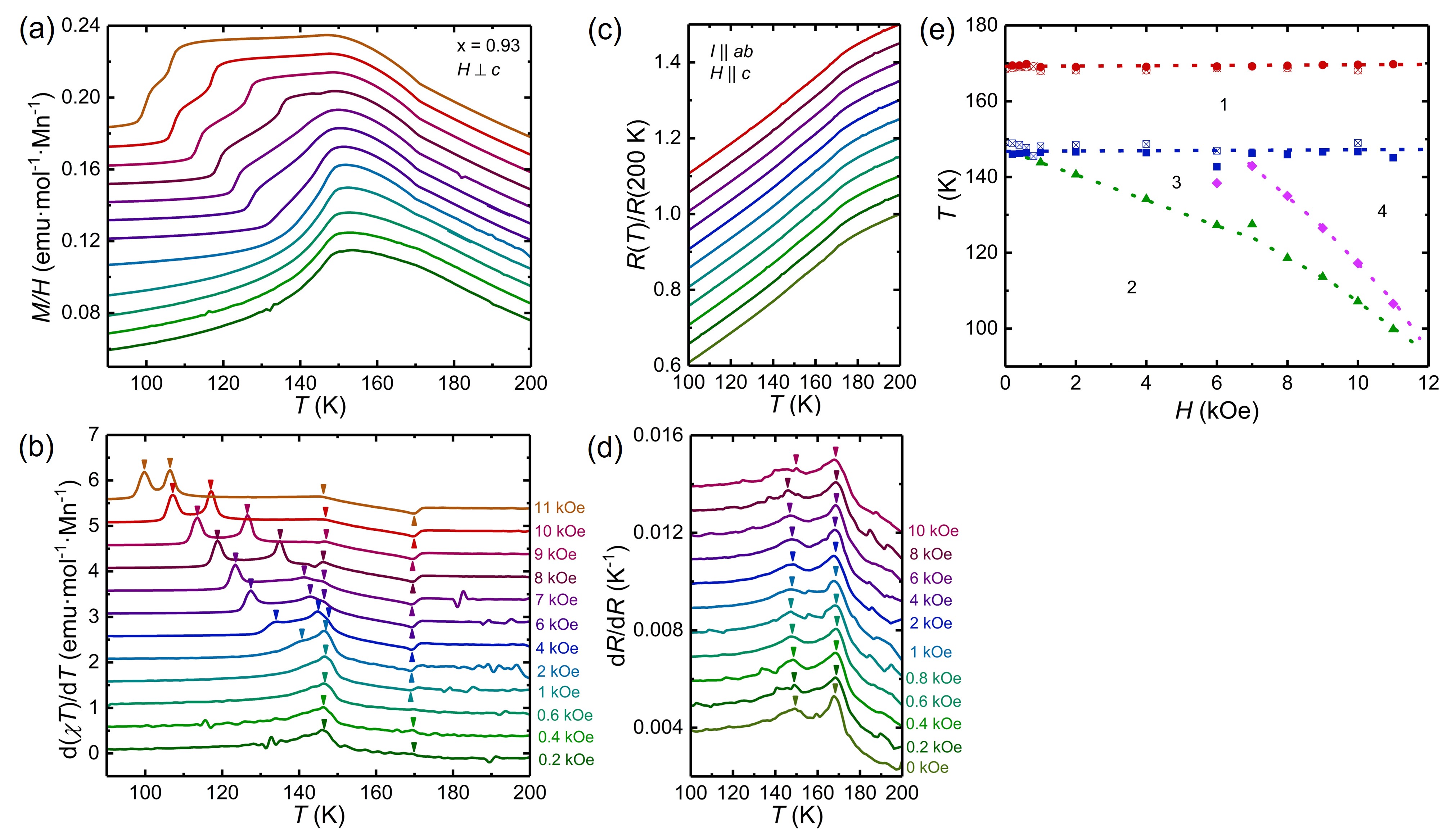}
    \caption[]{(a) and (c) show the temperature dependent \textit{M/H} and resistance data for a \textit{x} = 0.93 crystal and collected at different fields between 0.2--11 kOe. (b) and (d) display the derivatives, d($\chi$\textit{T})/d\textit{T} and d\textit{R}/d\textit{T}, of the curves in (a) and (c) respectively, with arrows marking the assigned magnetic transitions. (e) shows a \textit{H}-\textit{T} phase diagram assembled from the data in (a)--(d). The closed points are from the magnetic data and the crossed open points from the resistance data. All dashed lines are guides to the eye.}
    \label{MT_RT_H_x93}
\end{figure*}

\begin{figure*}[!t]
    \centering
    \includegraphics[width=0.5\linewidth]{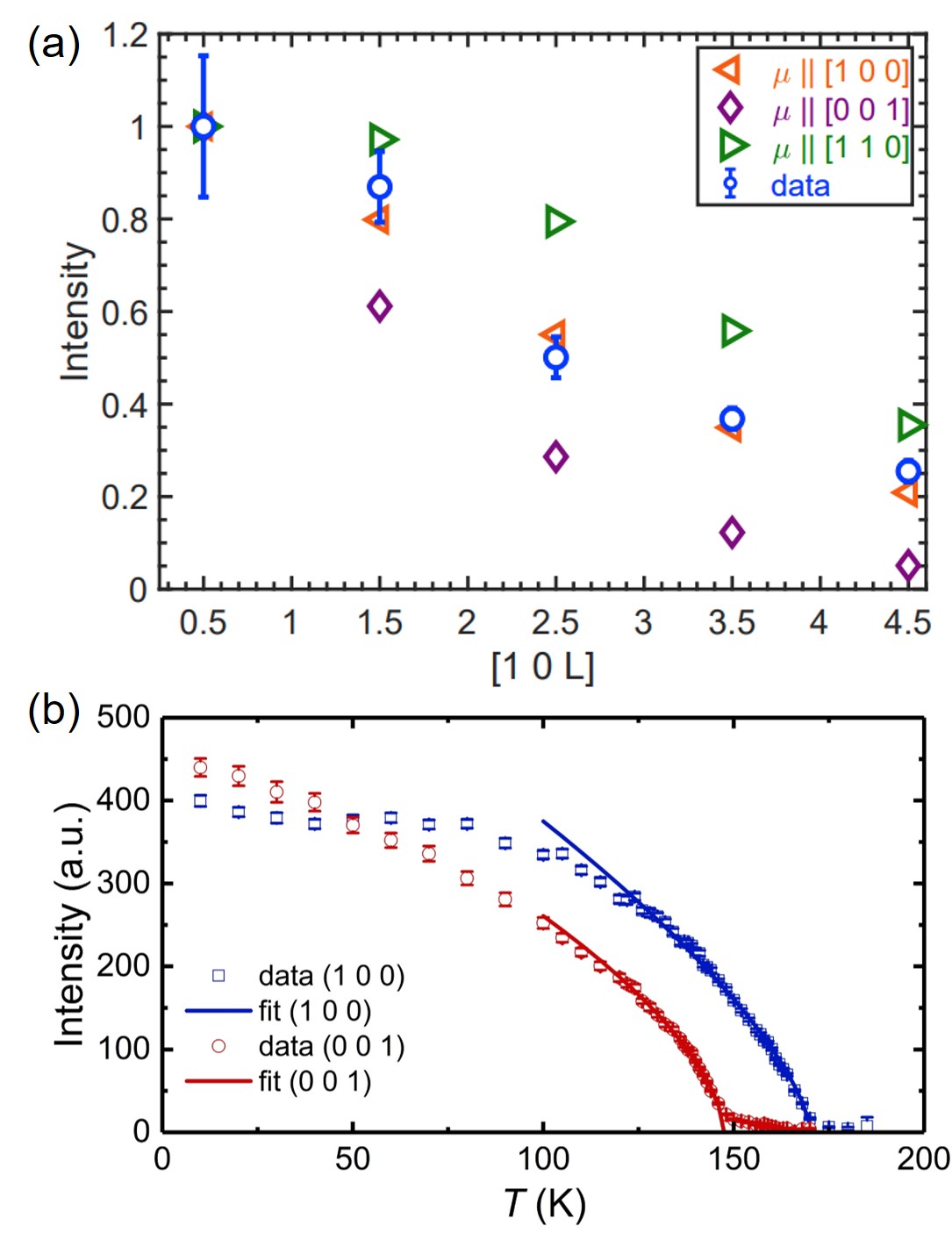}
    \caption[]{Results from single crystal neutron diffraction for an \textit{x} = 0.93 LaMn$_x$Sb$_2$ sample. (a) Intensity of the [1 0 L] Bragg peaks at 155 K. The blue points are the experimental data, and the green, purple, and orange liens the calculated intensities assuming the moments are respectively aligned along [1 0 0], [0 0 1], or [1 1 0]. Temperature dependence of the (1 0 0.5) and (0 0 1.5) Bragg peaks. The solid lines show the behavior expected for a mean-field like transition.}
    \label{SCneutron}
\end{figure*}

\subsection{Additional magnetic data for LaMn$_x$Sb$_2$}

The magnetic data presented in Figure \ref{MT_1kOe} in the main text shows a striking change in the behavior near \textit{x} = 0.89. The \textit{M/H} for \textit{x} = 0.89 prominently features a very strong, step-like drop at 117 K, indicating a first-order transition. The data for \textit{x} = 0.91 and \textit{x} = 0.92 samples show similar transitions (\textit{T}$_3$), but the drops in \textit{M/H} are not as sharp, making it ambiguous whether these are first- or second-order transitions. In our \textit{T}--\textit{x} phase diagram, shown in Figure \ref{T-x}, we interpret \textit{T}$_3$ as increasing in temperature as \textit{x} is lowered, such that \textit{T}$_3$ crosses above \textit{T}$_2$ near \textit{x} $\approx$ 0.9. Because such a crossing between \textit{T}$_2$ and \textit{T}$_3$ is only possible if one of the transition is first-order,\autocite{PhysRevB.43.2742} we show in Figure \ref{T3_hysteresis} a close-up view of \textit{T}$_3$ in \textit{x} = 0.91 and \textit{x} = 0.92 samples. Each dataset shows measurable hysteresis between warming and cooling sweeps, confirming \textit{T}$_3$ is a first-order transition, as expected.

The magnetic data for samples with \textit{x} $\geq$ 0.93 ambiguously suggest there are three magnetic transitions, the first occurring at $\approx$ 170-180 K and followed by two closely spaced transitions near 150 K, whereas the corresponding resistance data does not show any evidence for the lowest temperature transition. Figure \ref{MT_RT_H_x93}a and \ref{MT_RT_H_x93}b  show \textit{M/H} and resistance curves collected at fields ranging from 0.2--11 kOe for a \textit{x} = 0.93 sample, and the corresponding derivatives d($\chi$\textit{T})/d\textit{T} and d\textit{R}/d\textit{T} are given in Figures \ref{MT_RT_H_x93}b and \ref{MT_RT_H_x93}d. We find that in addition to the complex composition dependent magnetic properties outlined in Figures \ref{MT_1kOe} and \ref{T-x}, LaMn$_x$Sb$_2$ samples with \textit{x} $\geq$ 0.93 also have a rich \textit{H}-\textit{T} phase diagram as summarized in Figure \ref{MT_RT_H_x93}e. 

The 200 Oe data shows only two transitions, at 169 K and 146 K, consistent with the two transitions detected in the zero-field resistance data. The decrease in \textit{M/H} and peak in d($\chi$\textit{T})/d\textit{T} corresponding to the lower temperature transition both broaden as the field is increased, and by 2 kOe, has clearly split into two transitions at $\approx$ 146 K and 140 K. A fourth transition is discernible in the 4 kOe data, and the two lowest temperature transitions are steadily suppressed to lower temperatures as the field increases. These results indicate that the lowest temperature transitions suggested by the \textit{M/H} data in Figure \ref{MT_1kOe} for \textit{x} $\geq$ 0.93 correspond to a finite field phase and are therefore not included in \textit{T}-\textit{x} phase diagram. Whereas the two lower temperature transitions are not detectable in the resistance data, the higher temperature transitions are in good agreement with those determined from the magnetic data. This interpretation is consistent with our neutron diffraction data, which detect only two zero-field transitions for \textit{x} = 0.93. We finally emphasize that for samples with \textit{x} $<$ 0.93, \textit{M/H} measurements at lower and higher fields (up to 10 kOe) are not substantially different that the 1 kOe data in Figure \ref{MT_1kOe}.

\subsection{Additional neutron diffraction results}

The powder neutron diffraction presented in Figure \ref{neutron} of the main text implies a \textit{G}-type magnetic structure; however, we cannot confidently distinguish between arrangements in which the moments are aligned within the basal-plane or along the \textit{c}-axis. Analysis of single-crystal neutron diffraction measurements resolved the ambiguity. Figure \ref{SCneutron}a shows the intensities of the [1 0 L] Bragg peaks from single crystal neutron diffraction measurements that were conducted at 155 K. The dashed lines show the projected intensities assuming that the moments are aligned along [1 0 0], [0 0 1], or [1 1 0]. We find that the experimental data is in good agreement with the calculated results for moments along [1 0 0]; i.e. the single crystal neutron diffraction implies the Mn moments are aligned within the basal plane.

Figure \ref{SCneutron}b shows the temperature dependence of the (1 0 0.5) and (0 0 1.5) single crystal neutron diffraction peaks, indicating two transitions at $\approx$ 170 K and 150 K in good agreement with the \textit{M}(\textit{T}) and \textit{R}(\textit{T}) results. The solid lines show both transitions are mean-field like near \textit{T}$_N$.

\subsection{Magnetic data for Individual LaMn$_x$Sb$_2$ samples}

\begin{figure*}[!t]
    \centering
    \includegraphics[width=0.75\linewidth]{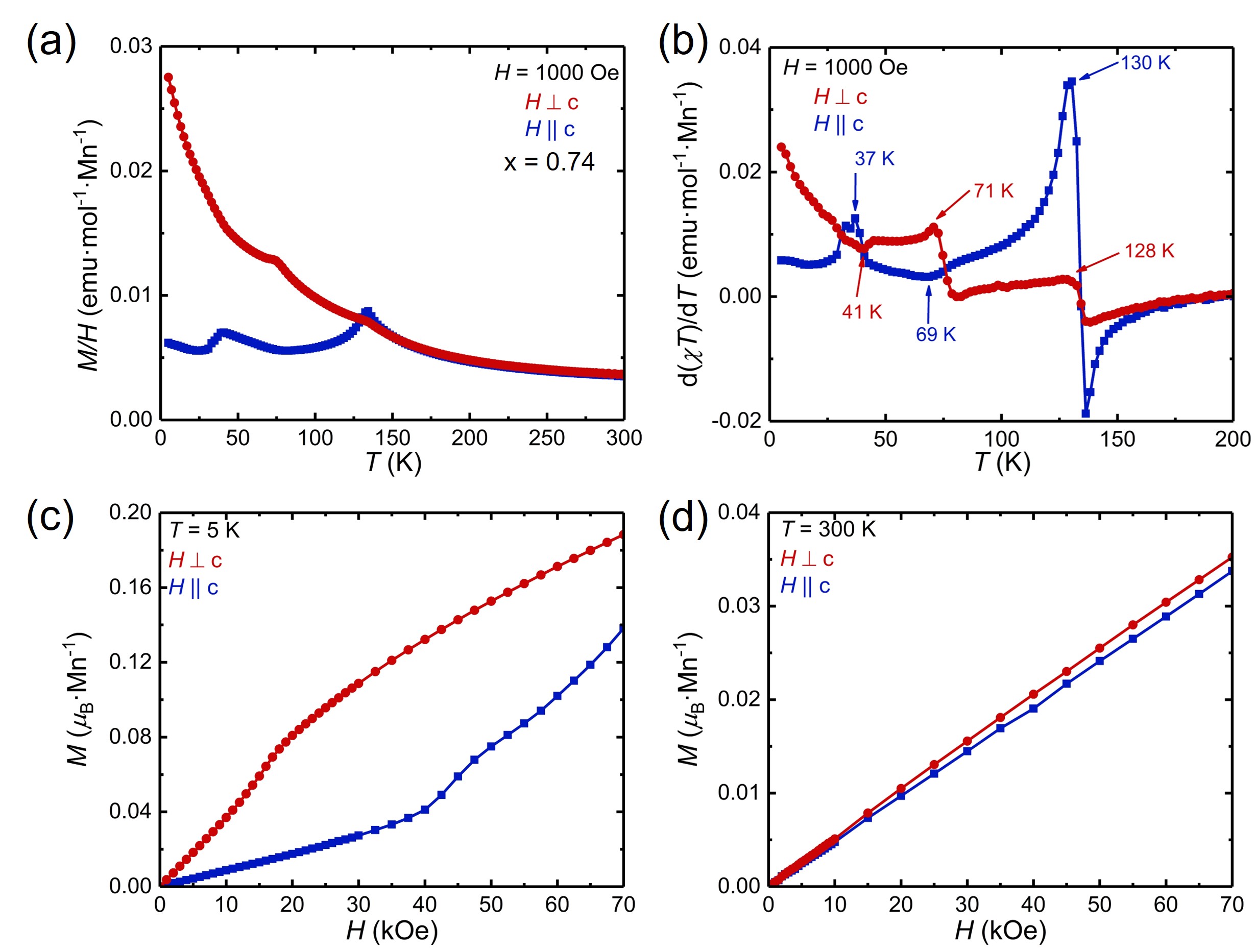}
    \caption[]{Magnetic properties of LaMn$_x$Sb$_2$ (x = 0.74) (a) Temperature dependent magnetic susceptibility, (b) temperature-susceptibility derivatives (d($\chi$\textit{T})/d\textit{T}) with the arrows marking the peaks used to assign magnetic transition temperatures, (c) field dependent magnetization at 5 K, and (d) field dependent magnetization at 300 K.}
    \label{data_x74}
\end{figure*}

\begin{figure*}[!b]
    \centering
    \includegraphics[width=0.75\linewidth]{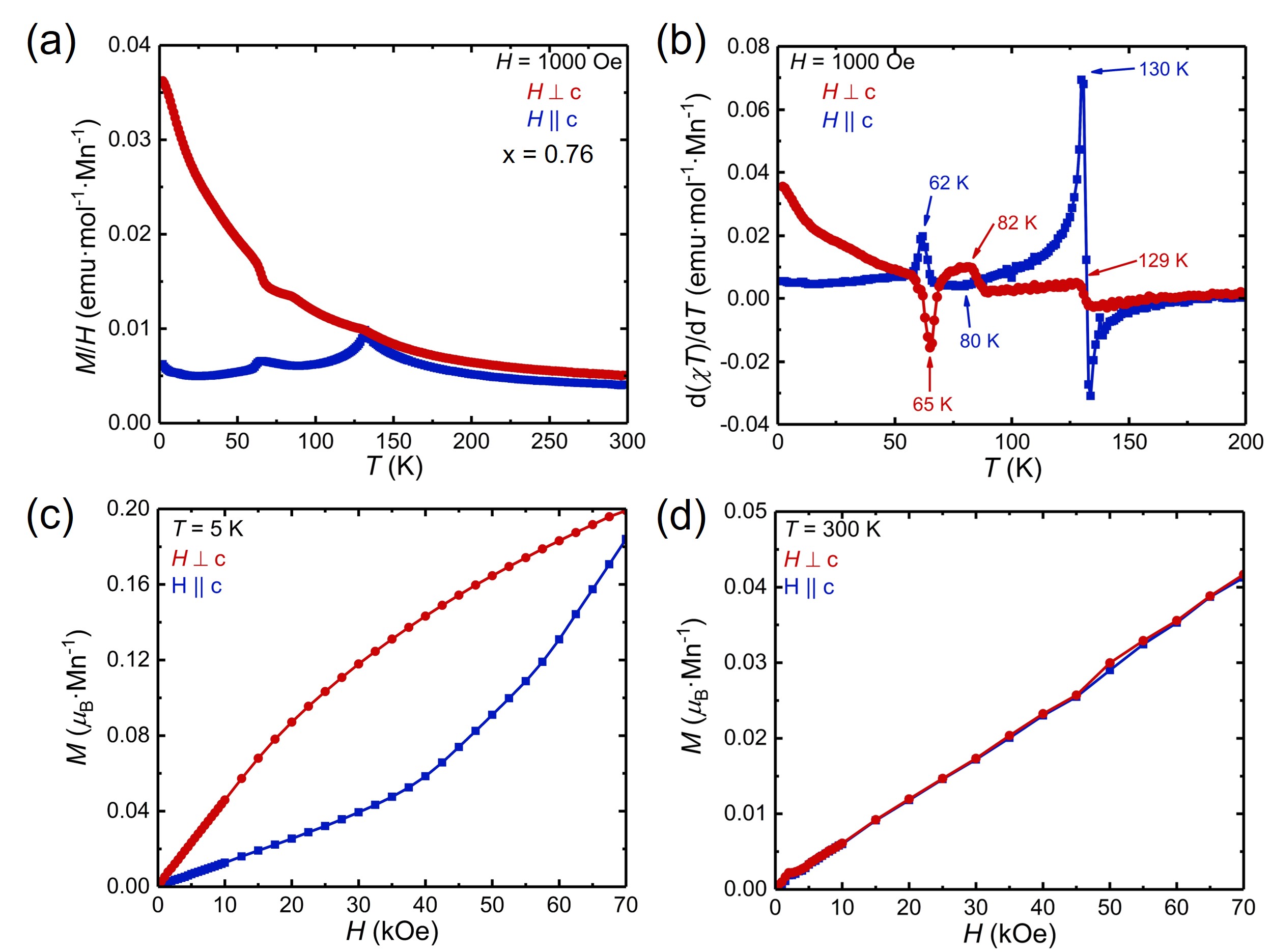}
    \caption[]{Magnetic properties of LaMn$_x$Sb$_2$ (x = 0.76) (a) Temperature dependent magnetic susceptibility, (b) temperature-susceptibility derivatives (d($\chi$\textit{T})/d\textit{T}) with the arrows marking the peaks used to assign magnetic transition temperatures, (c) field dependent magnetization at 5 K, and (d) field dependent magnetization at 300 K.}
    \label{data_x76}
\end{figure*}

\begin{figure*}[t!]
    \centering
    \includegraphics[width=0.75\linewidth]{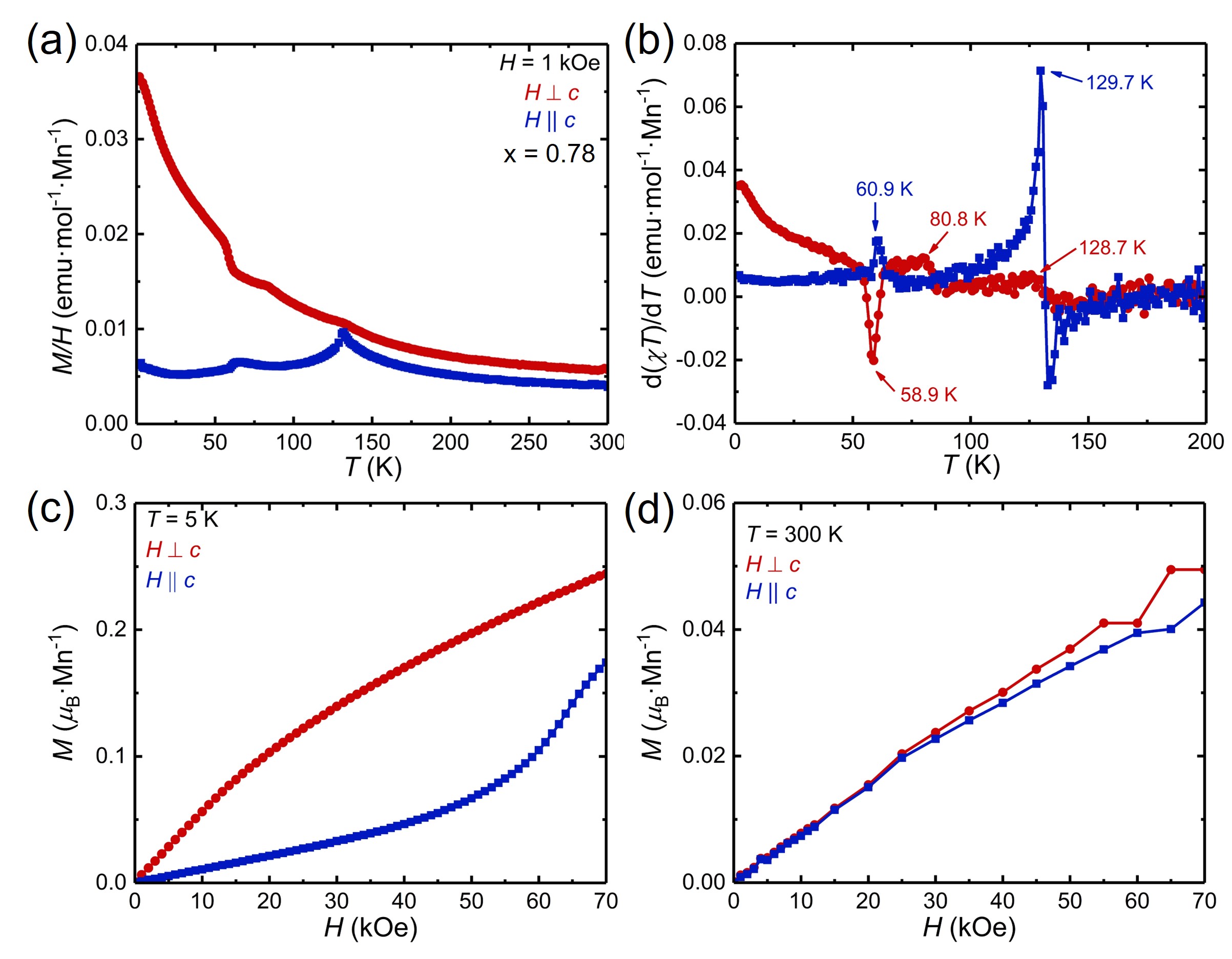}
    \caption[]{Magnetic properties of LaMn$_x$Sb$_2$ (x = 0.78) (a) Temperature dependent magnetic susceptibility, (b) temperature-susceptibility derivatives (d($\chi$\textit{T})/d\textit{T}) with the arrows marking the peaks used to assign magnetic transition temperatures, (c) field dependent magnetization at 5 K, and (d) field dependent magnetization at 300 K.}
    \label{data_x78}
\end{figure*}

\begin{figure*}[b!]
    \centering
    \includegraphics[width=0.70\linewidth]{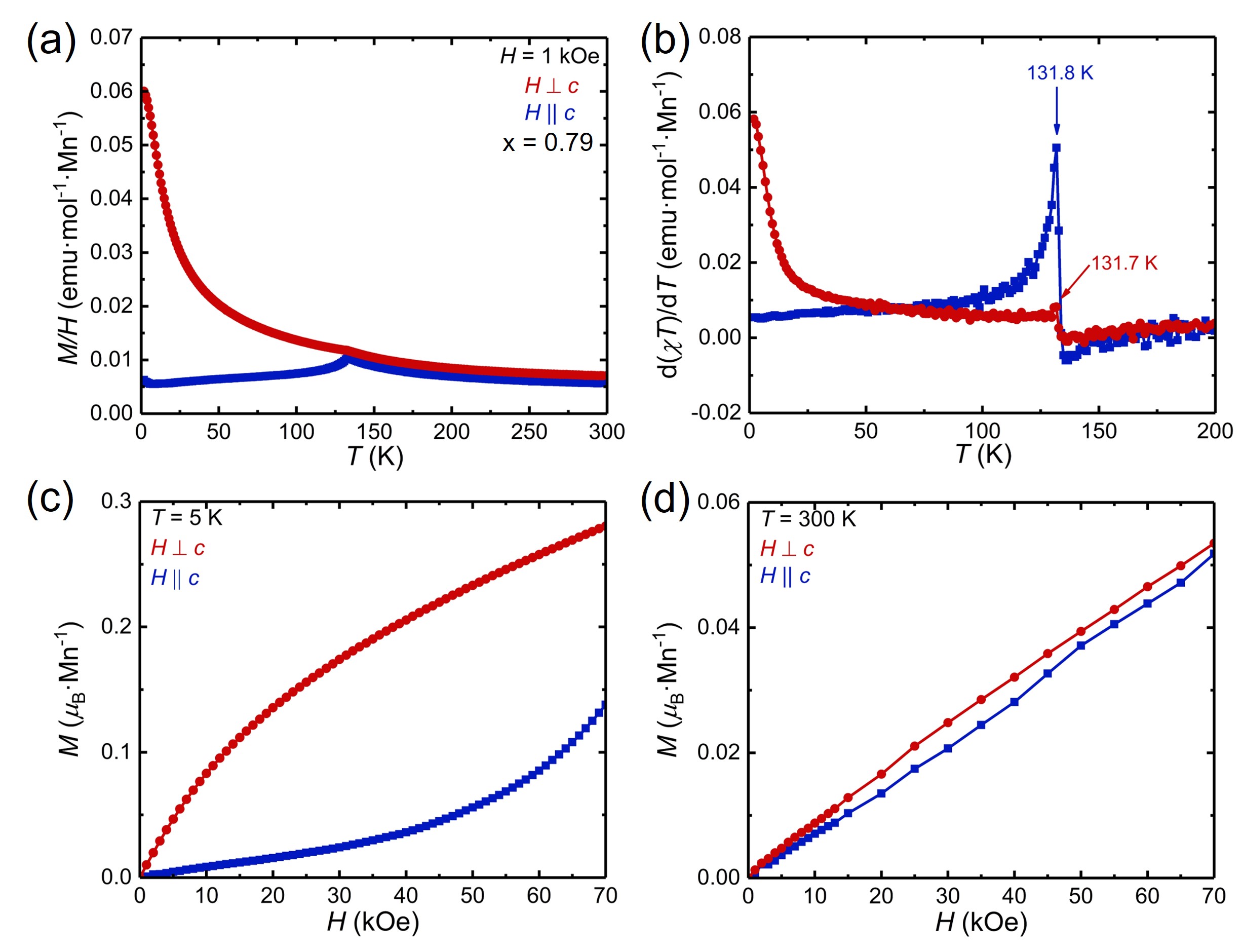}
    \caption[]{Magnetic properties of LaMn$_x$Sb$_2$ (x = 0.79) (a) Temperature dependent magnetic susceptibility, (b) temperature-susceptibility derivatives (d($\chi$\textit{T})/d\textit{T}) with the arrows marking the peaks used to assign magnetic transition temperatures, (c) field dependent magnetization at 5 K, and (d) field dependent magnetization at 300 K.}
    \label{data_x79}
\end{figure*}

\begin{figure*}[!t]
    \centering
    \includegraphics[width=0.70\linewidth]{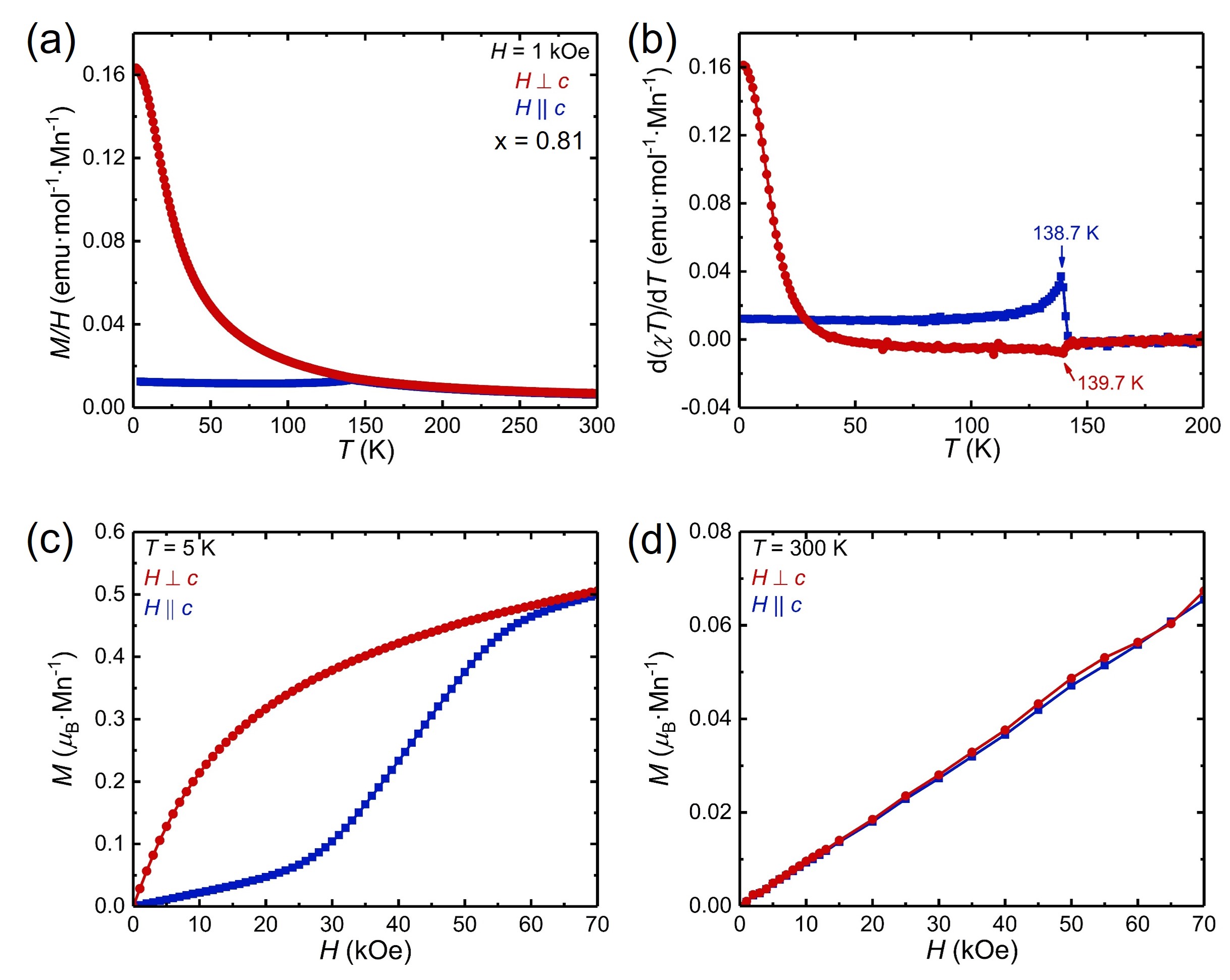}
    \caption[]{Magnetic properties of LaMn$_x$Sb$_2$ (x = 0.81) (a) Temperature dependent magnetic susceptibility, (b) temperature-susceptibility derivatives (d($\chi$\textit{T})/d\textit{T}) with the arrows marking the peaks used to assign magnetic transition temperatures, (c) field dependent magnetization at 5 K, and (d) field dependent magnetization at 300 K.}
    \label{data_x81}
\end{figure*}

\begin{figure*}[!b]
    \centering
    \includegraphics[width=0.72\linewidth]{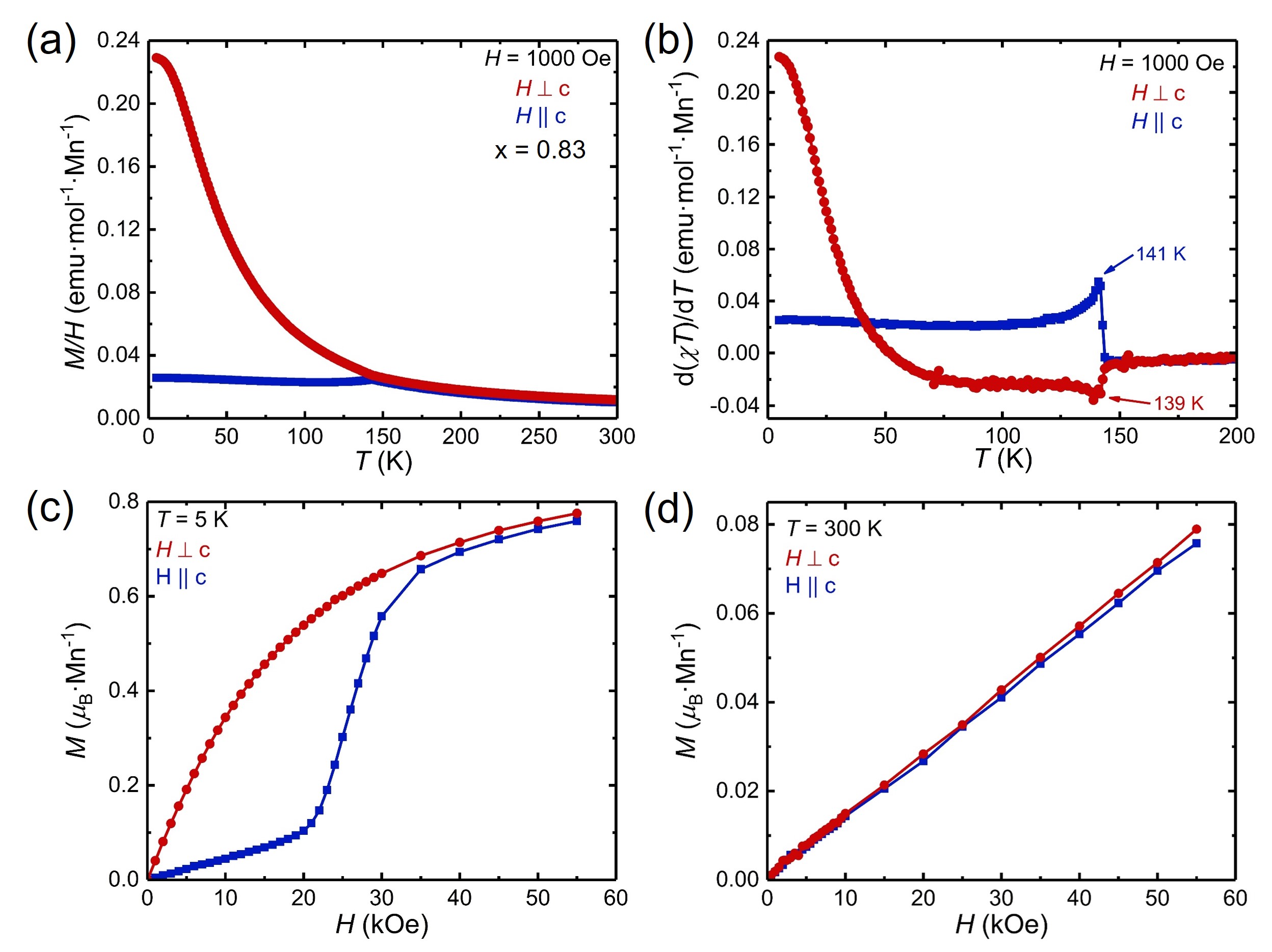}
    \caption[]{Magnetic properties of LaMn$_x$Sb$_2$ (x = 0.83) (a) Temperature dependent magnetic susceptibility, (b) temperature-susceptibility derivatives (d($\chi$\textit{T})/d\textit{T}) with the arrows marking the peaks used to assign magnetic transition temperatures, (c) field dependent magnetization at 5 K, and (d) field dependent magnetization at 300 K.}
    \label{data_x83}
\end{figure*}

\begin{figure*}[!t]
    \centering
    \includegraphics[width=0.72\linewidth]{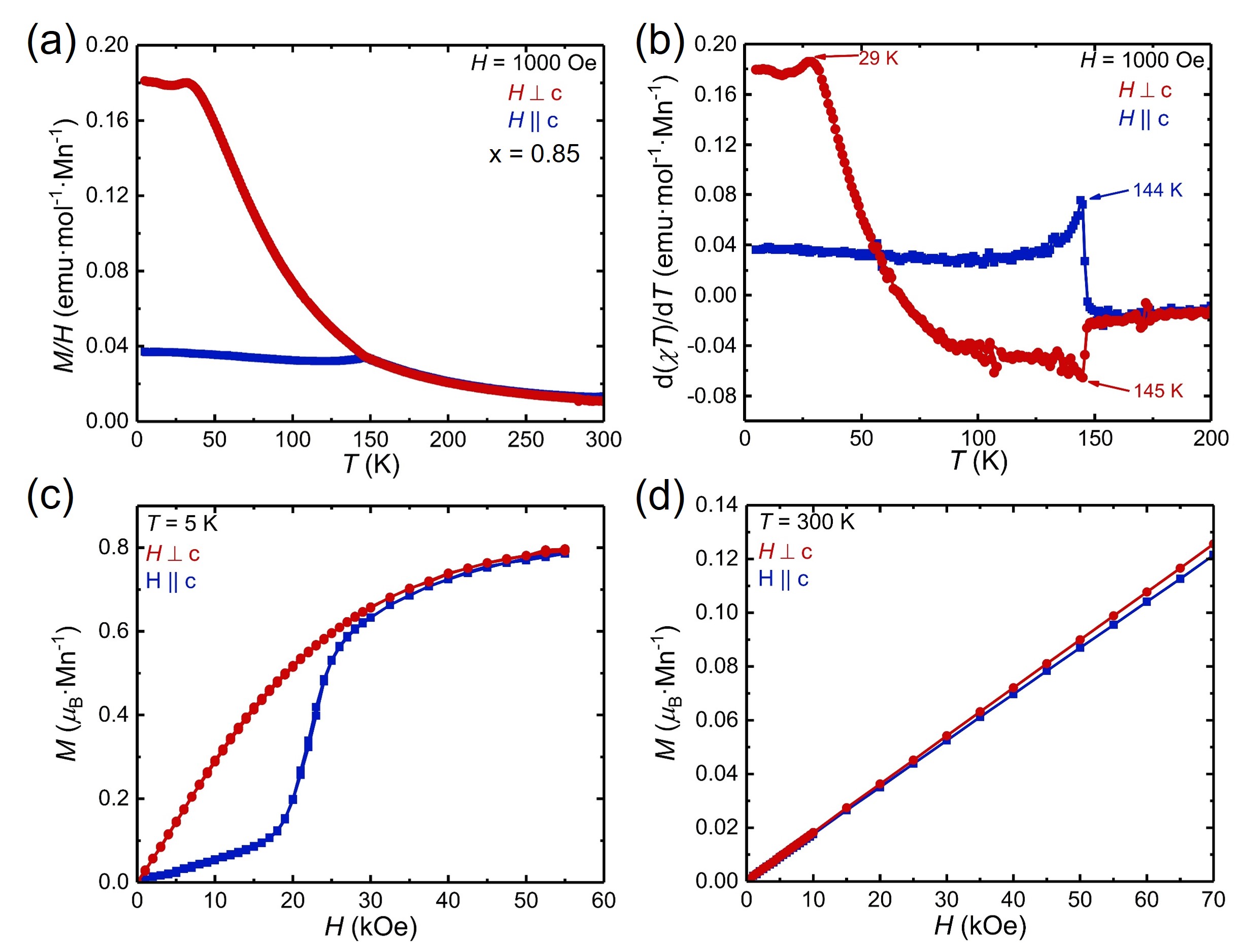}
    \caption[]{Magnetic properties of LaMn$_x$Sb$_2$ (x = 0.85) (a) Temperature dependent magnetic susceptibility, (b) temperature-susceptibility derivatives (d($\chi$\textit{T})/d\textit{T}) with the arrows marking the peaks used to assign magnetic transition temperatures, (c) field dependent magnetization at 5 K, and (d) field dependent magnetization at 300 K.}
    \label{data_x85}
\end{figure*}

\begin{figure*}[!b]
    \centering
    \includegraphics[width=0.72\linewidth]{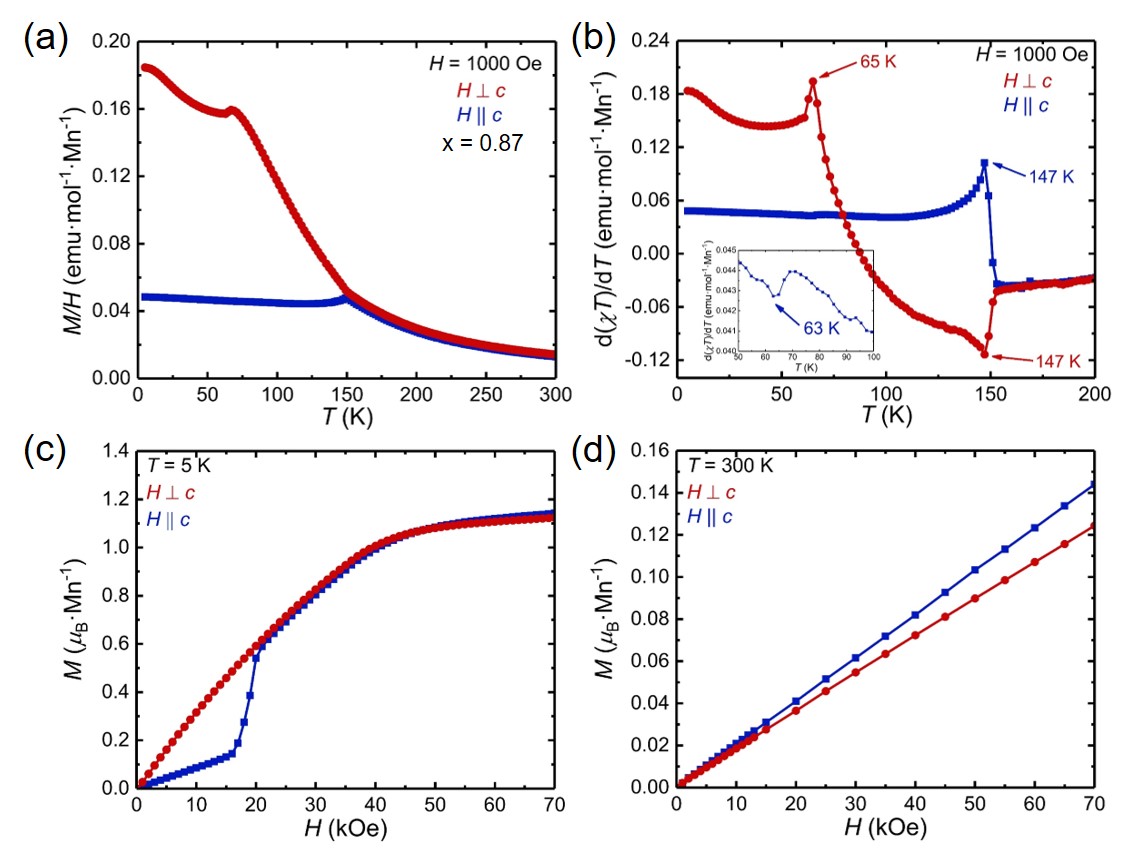}
    \caption[]{Magnetic properties of LaMn$_x$Sb$_2$ (x = 0.87) (a) Temperature dependent magnetic susceptibility, (b) temperature-susceptibility derivatives (d($\chi$\textit{T})/d\textit{T}) with the arrows marking the peaks used to assign magnetic transition temperatures. The inset shows a closeup of the low temperature transition in the \textit{H} $\parallel$ \textit{c} orientation. (c) Field dependent magnetization at 5 K, and (d) field dependent magnetization at 300 K.}
    \label{data_x86}
\end{figure*}

\begin{figure*}[!t]
    \centering
    \includegraphics[width=0.72\linewidth]{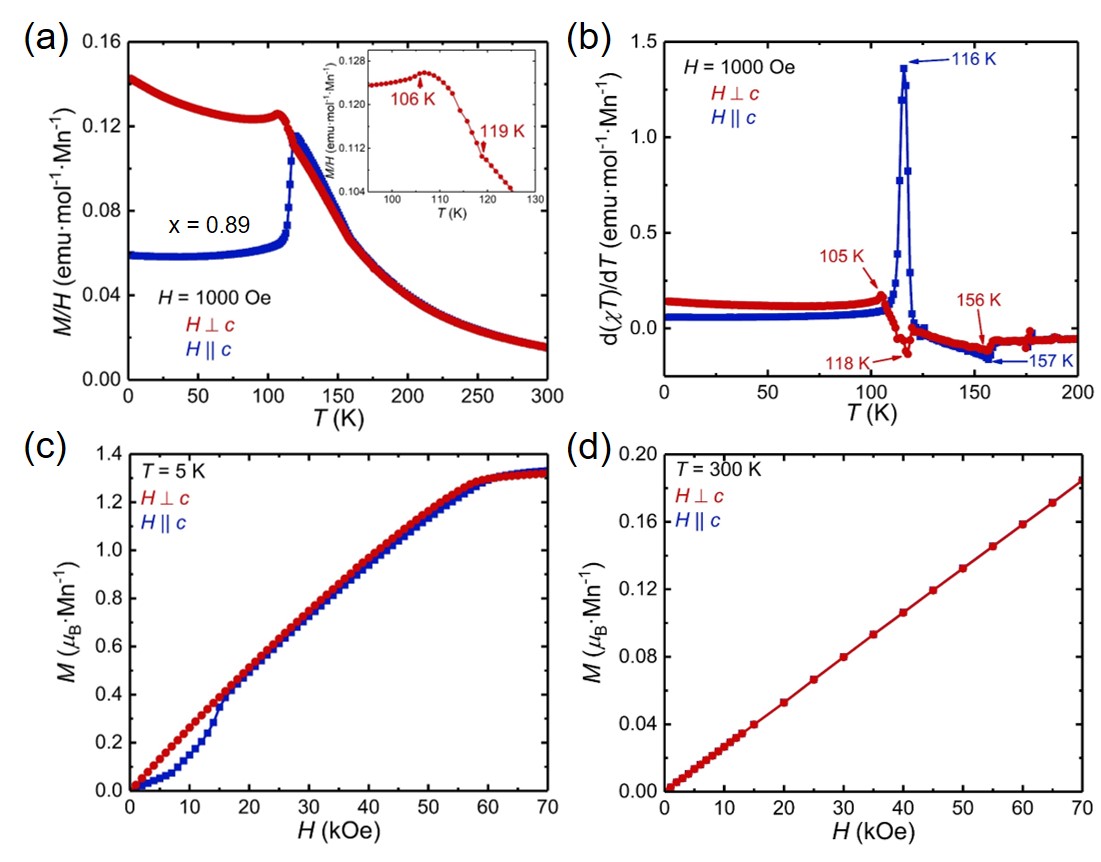}
    \caption[]{Magnetic properties of LaMn$_x$Sb$_2$ (x = 0.89) (a) Temperature dependent magnetic susceptibility, (b) temperature-susceptibility derivatives (d($\chi$\textit{T})/d\textit{T}) with the arrows marking the peaks used to assign magnetic transition temperatures, (c) field dependent magnetization at 5 K, and (d) field dependent magnetization at 300 K. The inset in (a) shows a close up view of the \textit{M/H} for \textit{H} $\perp$ \textit{c} with the two lower temperature transitions marked with arrows.}
    \label{data_x89}
\end{figure*}

\begin{figure*}[!b]
    \centering
    \includegraphics[width=0.72\linewidth]{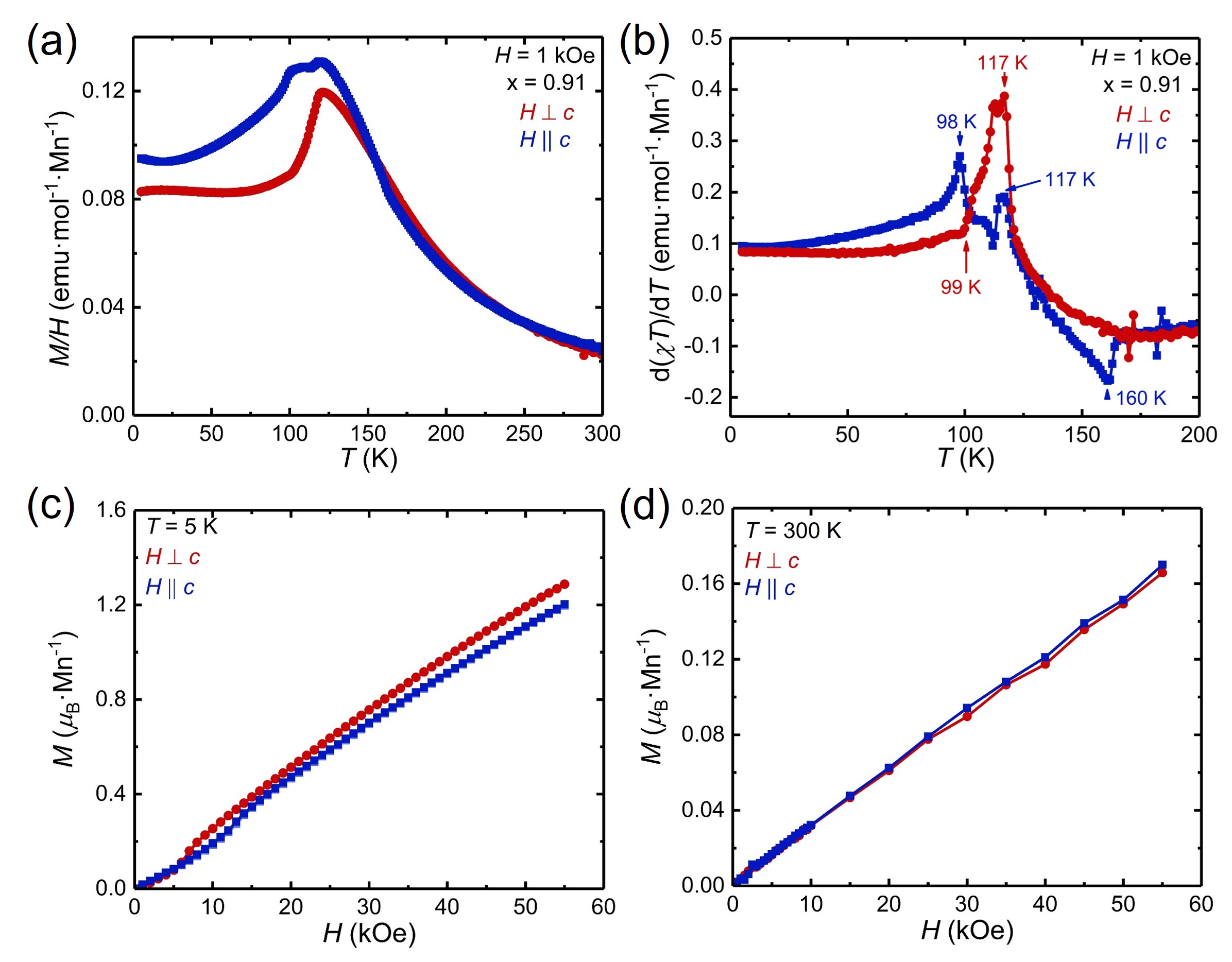}
    \caption[]{Magnetic properties of LaMn$_x$Sb$_2$ (x = 0.91) (a) Temperature dependent magnetic susceptibility, (b) temperature-susceptibility derivatives (d($\chi$\textit{T})/d\textit{T}) with the arrows marking the peaks used to assign magnetic transition temperatures, (c) field dependent magnetization at 5 K, and (d) field dependent magnetization at 300 K.}
    \label{data_x91}
\end{figure*}

\begin{figure*}[!b]
    \centering
    \includegraphics[width=0.72\linewidth]{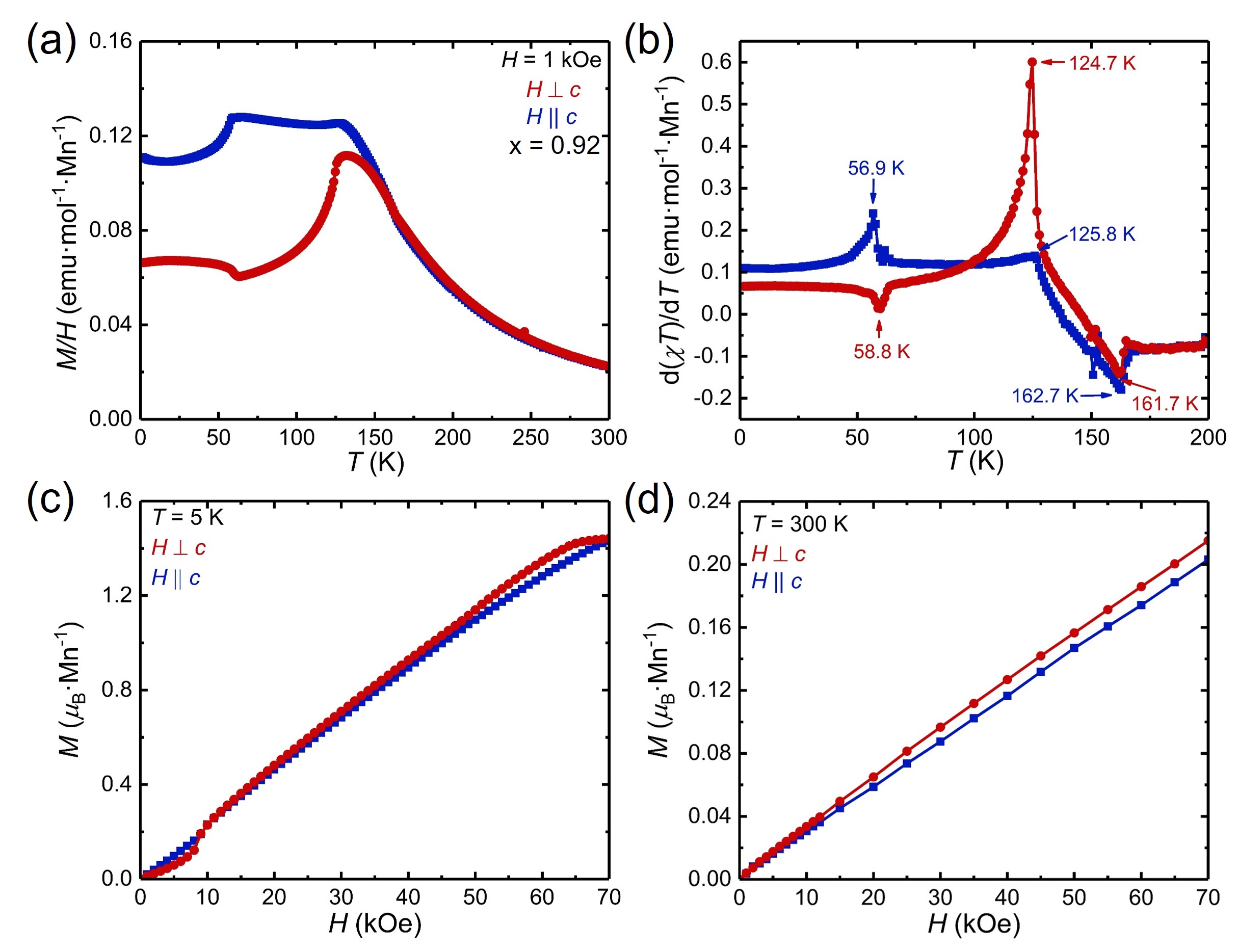}
    \caption[]{Magnetic properties of LaMn$_x$Sb$_2$ (x = 0.92) (a) Temperature dependent magnetic susceptibility, (b) temperature-susceptibility derivatives (d($\chi$\textit{T})/d\textit{T}) with the arrows marking the peaks used to assign magnetic transition temperatures, (c) field dependent magnetization at 5 K, and (d) field dependent magnetization at 300 K.}
    \label{data_x92}
\end{figure*}

\begin{figure*}[!t]
    \centering
    \includegraphics[width=0.72\linewidth]{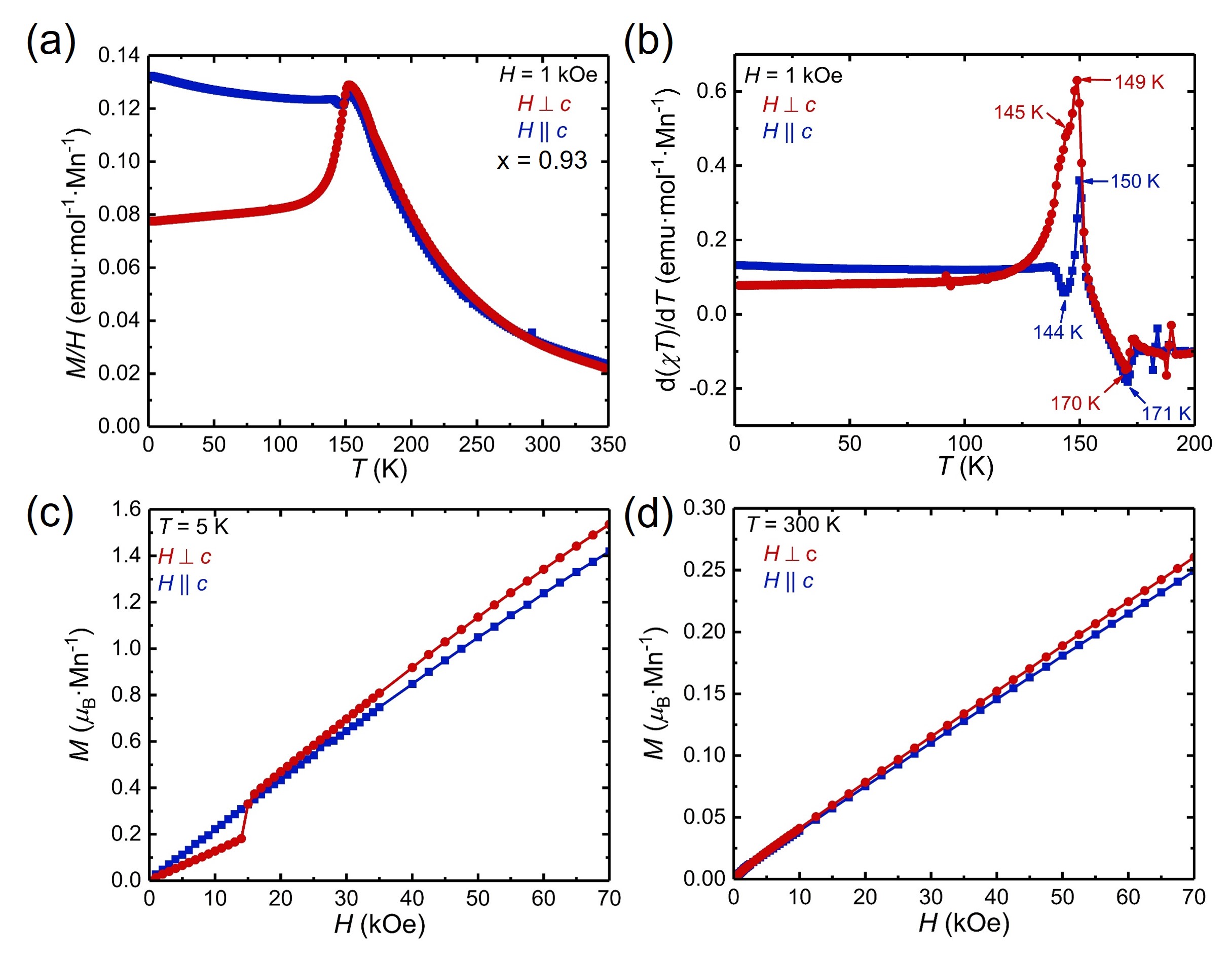}
    \caption[]{Magnetic properties of LaMn$_x$Sb$_2$ (x = 0.93) (a) Temperature dependent magnetic susceptibility, (b) temperature-susceptibility derivatives (d($\chi$\textit{T})/d\textit{T}) with the arrows marking the peaks used to assign magnetic transition temperatures, (c) field dependent magnetization at 5 K, and (d) field dependent magnetization at 300 K.}
    \label{data_x93}
\end{figure*}

\begin{figure*}[!b]
    \centering
    \includegraphics[width=0.72\linewidth]{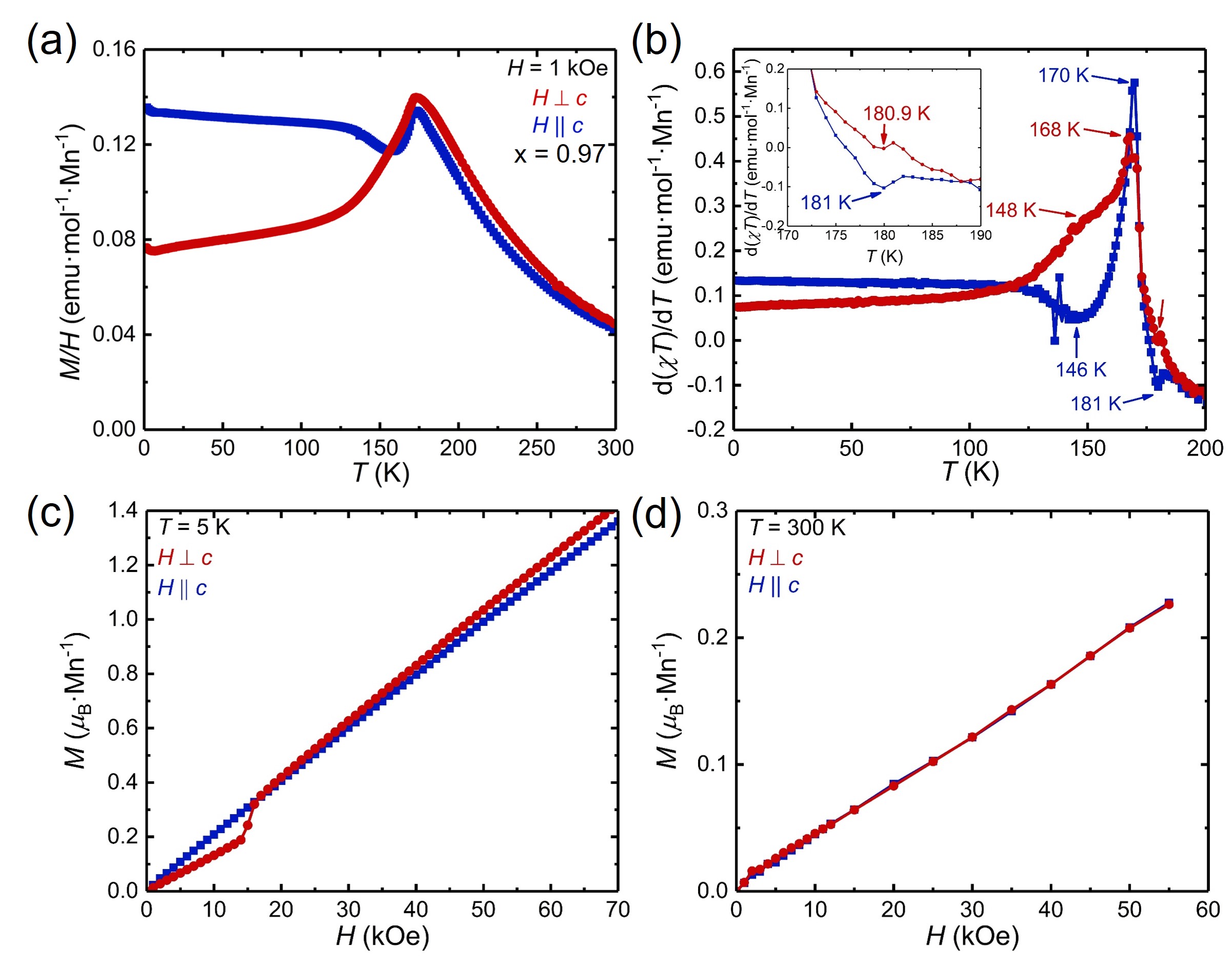}
    \caption[]{Magnetic properties of LaMn$_x$Sb$_2$ (x = 0.97) (a) Temperature dependent magnetic susceptibility, (b) temperature-susceptibility derivatives (d($\chi$\textit{T})/d\textit{T}) with the arrows marking the peaks used to assign magnetic transition temperatures. The inset shows a closeup of the transition near 180 K. (c) Field dependent magnetization at 5 K, and (d) field dependent magnetization at 300 K.}
    \label{data_x97}
\end{figure*}

For clarity, Figures \ref{data_x74}--\ref{data_x97} shown below give the individual \textit{M}(\textit{T})/\textit{H}, d($\chi$\textit{T})/d\textit{T}, and \textit{M}(\textit{H}) data for each sample of LaMn$_x$Sb$_2$ separated for unique values of \textit{x}. The data is the same as that presented in Figures \ref{MT_1kOe}, \ref{MH_5K}, and Figure \ref{MH_300K}.

\newpage

\end{singlespace}

\end{document}